\documentclass[reprint,showpacs,groupedaddress,floatfix,prb,superscriptaddress]{revtex4-1}

\usepackage{amssymb,amsmath}
\usepackage{graphicx}
\usepackage{float}

\usepackage{hyperref} 
\hypersetup{
     colorlinks   = true,
     citecolor    = blue,
     linkcolor   = blue
}
\graphicspath{{./Images/}}

\usepackage{siunitx}

\newcommand{\0}{\phantom{0}}

\newcommand{\2}{\phantom{$(0)$}}
\newcommand{\+}{\phantom{+}}
\newcommand{\ms}{\mspace{-4mu}}
\newcommand{\TO}{\text{TO}}
\newcommand{\LO}{\text{LO}}

\newcommand{\ie}{i.\,e.}

\begin{document}

\title{CZTS Raman spectra beyond kesterite: a first-principles study.}

\author{S. P. Ramkumar}
\affiliation{IMCN-MODL, Universit\'{e} catholique de Louvain, Chemin des \'{E}toiles 8, B-1348 Louvain-la-Neuve, Belgium}
\affiliation{Department of Materials Science and Engineering, University of California, Merced, Merced, California 95343, USA}
\affiliation{European Theoretical Spectroscopy Facility (ETSF)}

\author{H. P. C. Miranda}
\affiliation{IMCN-MODL, Universit\'{e} catholique de Louvain, Chemin des \'{E}toiles 8, B-1348 Louvain-la-Neuve, Belgium}
\affiliation{European Theoretical Spectroscopy Facility (ETSF)}

\author{X. Gonze}
\affiliation{IMCN-MODL, Universit\'{e} catholique de Louvain, Chemin des \'{E}toiles 8, B-1348 Louvain-la-Neuve, Belgium}
\affiliation{European Theoretical Spectroscopy Facility (ETSF)}
\affiliation{Skolkovo Institute of Science and Technology, Skolkovo Innovation Center, Nobel St. 3, Moscow, 143026, Russia}

\author{G.-M. Rignanese}
\affiliation{IMCN-MODL, Universit\'{e} catholique de Louvain, Chemin des \'{E}toiles 8, B-1348 Louvain-la-Neuve, Belgium}
\affiliation{European Theoretical Spectroscopy Facility (ETSF)}


\begin{abstract}
Cu$_2$ZnSnS$_4$ is an earth-abundant photovoltaic absorber material predicted to provide a sustainable solution for commercial solar applications. One of the main limitations restricting its commercialization is the issue of cation disorder. 
Raman spectroscopy has been a sought after technique to characterize disorder in CZTS while a clear consensus between theoretical and experimental results is yet to be achieved. In the present study, via the virtual crystal approximation, we take into account the progressive nature of Cu-Zn disorder in CZTS: we obtain the phonon frequencies at zone-center within the density functional perturbation theory formalism, and further compute the Raman spectra for the disordered phases, achieving a consensus between theory and experiment. 
These calculations confirm the presence of complete disorder in Cu-Zn 2$a$, 2$c$ and 2$d$ Wyckoff sites. 
They also show that the Raman intensities of two prominent $A$ phonon modes 
characterized by motion of S atoms, also known to be experimentally significant, play a key role in understanding the nature of disorder in CZTS.
\end{abstract}

\maketitle

\section{Introduction}
\label{sec:intro}

\begin{figure*}[t]
\centering
\includegraphics{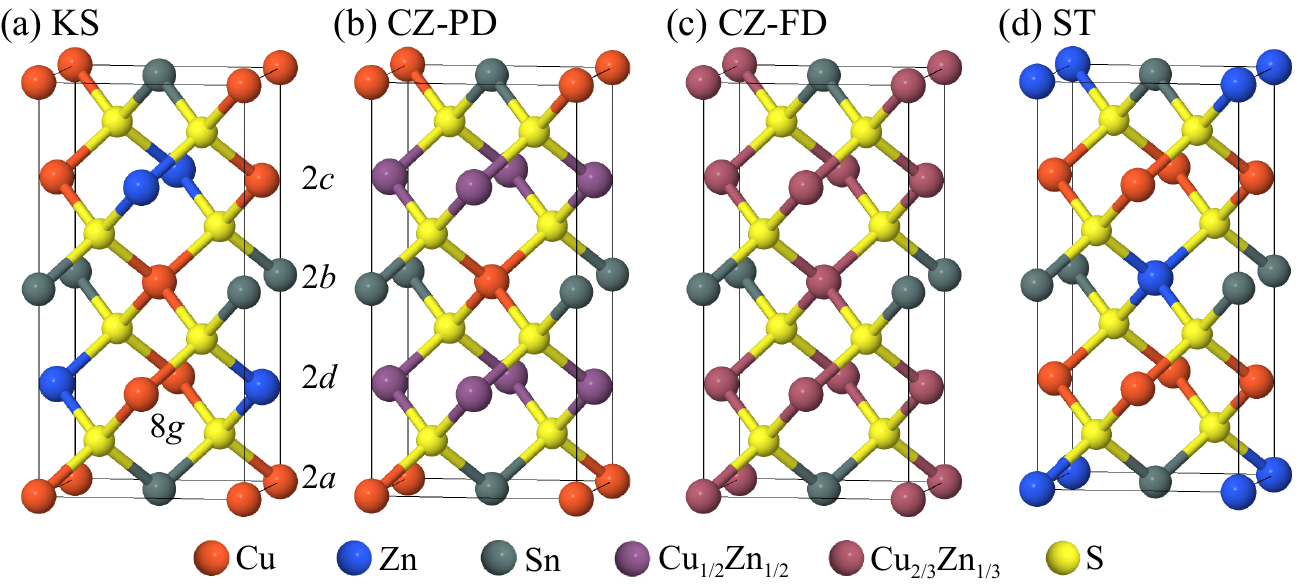}
\caption{Tetragonal unit cells of CZTS: (a) kesterite (KS), (b) Cu-Zn plane disordered (CZ-PD), (c) Cu-Zn fully disordered (CZ-FD), and (d) stannite (ST).
}
\label{fig:unit-cells}
\end{figure*}

In light of climate change due to greenhouse gas emissions, photovoltaics (PV) is expected to play an important role to meet the growing energy demand on a sustainable basis~\cite{breyer2017role}. Si solar cells currently dominate the PV market, but the more versatile thin-film technology (mainly, based on CdTe and CuIn$_x$Ga$_{1-x}$(S,Se)$_2$) has almost matched the former in performance, reaching efficiencies close to 21\%\cite{green2017solar}.
Nonetheless, almost all of the present commercially available PV technologies suffer from various drawbacks (be it the amount of energy needed to fabricate them, the low availability or the toxicity of one or more constituting elements) which will likely limit their role in large-scale applications.
To counter such drawbacks, PV devices based on Cu$_2$ZnSnS$_4$ (CZTS) with earth abundant and non-toxic elements have garnered tremendous interest in the last decade.
As a downside, their efficiency is only close to 8.6\%\cite{shin2013thin}, making them commercially not viable at the moment.
One major reason for the low efficiency in CZTS has been attributed to the presence of Cu-Zn disorder, confirmed by several experimental and theoretical studies, including X-ray and neutron diffraction~\cite{schorr2011crystal}, Raman spectra~\cite{scragg2014low,skelton2015vibrational,ramkumar2016first,dimitrievska2017structural,dimitrievska2017structural}, nuclear magnetic resonance (NMR)~\cite{paris2015119sn,paris2014solid}, and photoluminescence measurements~\cite{scragg2016cu}.  

Experimentally, CZTS is found to crystallize in the kesterite (KS) structure (with space group $I\bar{4}$)~\cite{schorr2007neutron,schorr2011crystal,lafond2014x}, illustrated in Fig.~\ref{fig:unit-cells}(a).
It has alternating Cu/Zn and Cu/Sn planes along the [001] direction and four different cationic sites: one Cu is located at the Wyckoff position 2$a$ (0,0,0), the other Cu at 2$c$ (0,1{}/2,1{}/4), Zn at 2$d$ (0,1{}/2,3{}/4) and Sn at 2$b$ (0,0,1{}/2).
Until recently, the generally accepted picture was that, above the critical temperature $T_c$ (ranging between 533 and 552~K~\cite{scragg2014low,ritscher2016order}), KS evolves to CZ-PD which is characterized by a disorder in the Cu-Zn planes (Fig. 1(b)).
However, a high-resolution neutron diffraction experiment~\cite{bosson2017cation}, soon followed by theoretical works relying on density functional theory (DFT)~\cite{ramkumar2018insights,wallace2019atomistic,zheng2019cu}, showed that it actually goes to CZ-FD, a phase in which all Cu-Zn sites are fully disordered (Fig.~\ref{fig:unit-cells}(c)). 

\begin{figure}[h]
\centering
\includegraphics{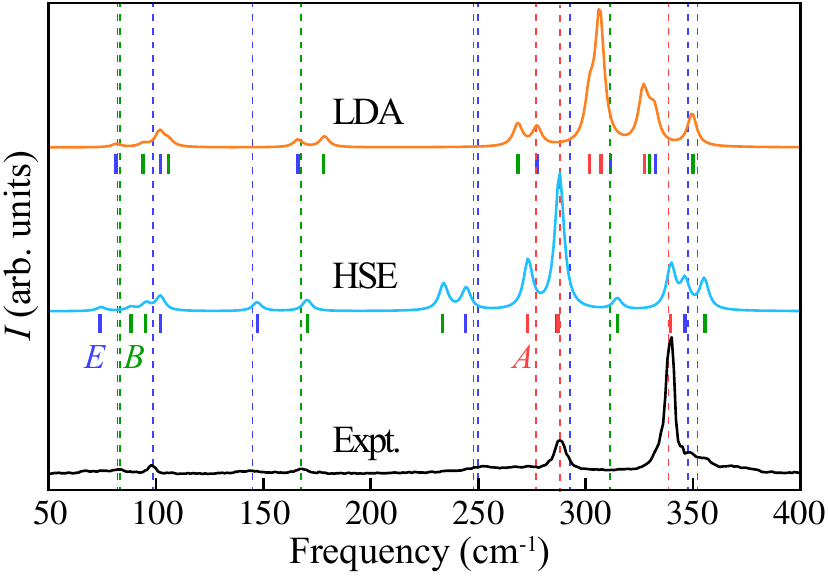}
\caption{Raman spectra of CZTS. Upper and middle curves show the Raman spectrum for the KS structure, computed using LDA and HSE (using LDA intensities) TO modes. 
For comparison with experiment~\cite{dimitrievska2014multiwavelength} (black curve in the bottom), a temperature of 300~K is chosen for the Bose-Einstein occupation factors. 
The red, green and blue dotted lines denote the TO frequencies of $A$, $B$, and $E$ modes obtained from polarized Raman measurements of~\citet{guc2016polarized}.  
}
\label{fig:raman-spectra}
\end{figure}

While a good agreement has been reached between the theoretical and experimental observations in terms of phase characteristics and partial occupancy, this is not completely the case for the vibrational properties and, in particular, for Raman spectroscopy which has been widely used to try to characterize CZTS~\cite{dimitrievska2017structural,dimitrievska2014multiwavelength,fontane2012vibrational,paris2014solid,cheng2011imaging,khare2012calculation}.
Experimental Raman spectra~\cite{fontane2012vibrational,dimitrievska2017structural,dimitrievska2014multiwavelength,paris2014solid,cheng2011imaging,khare2012calculation} typically show two prominent peaks at 287 and 331~cm$^{-1}$ with the second one displaying a higher intensity (see Fig.~\ref{fig:raman-spectra}).

From the theoretical standpoint, all previous theoretical works~\cite{khare2012calculation, gurel2011characterization,skelton2015vibrational,ramkumar2016first,dimitrievska2017structural,monserrat2018role} have focused on ordered structures.
Most of them relied on semi-local exchange-correlation (XC) functionals such as the local density approximation (LDA) or the generalized gradient approximation (GGA).
In this framework, none of the investigated phases was found to present a spectrum corresponding to the experimental one within the error bar that is usually expected.
Typically, the higher frequency phonon modes, which are predominantly dominated by the motion of S atoms, show a red shift of 15~cm$^{-1}$ with respect to experiment.
Furthermore, in the two earlier works~\cite{ramkumar2016first,skelton2015vibrational} in which the Raman intensities were calculated, the first prominent peak was found to be higher in intensity than the second one, in clear contrast with experimental findings.

Recently, \citet{monserrat2018role} resolved one of these discrepancies, improving the agreement in terms of the phonon frequencies by using a hybrid XC functional which better accounts for the polarisabilities of the S atoms.
However, they did not compute the Raman intensities and focused only on the phonon density of states.
So the question is still open whether the remaining intensity discrepancy is to be ascribed to the XC functional or to the presence of disorder in the structure.

In order to address this issue, we perform first-principles calculations based on density functional theory (DFT)~\cite{hohenberg1964inhomogeneous} and density functional perturbation theory (DFPT)~\cite{giannozzi1991ab,gonze1997dynamical,gonze1997first} adopting the virtual crystal approximation (VCA)~\cite{jaros1985electronic,bellaiche2000virtual} to account for the disorder.
We show that the changes in Raman intensity resulting from the disorder are sufficient to reproduce the experimental intensities.

The paper is organized as follows.
Sec.~\ref{sec:computational-details} describes the computational methods and the different VCA structures considered in the study.
In Sec.~\ref{sec:results-discussion}, our main results are presented and discussed.
Sec.~\ref{sec:phonons} shows a comparison of the calculated phonon frequencies with experiments. In particular, the effect of exchange-correlation functionals (semi-local vs. hybrid) is discussed. In Sec.~\ref{sec:raman-spectra}, the calculated Raman spectra are analyzed for all the phases, identifying their key features. Finally, Sec.~\ref{sec:amodes-evolution} focuses on the evolution of the different $A$ modes (Raman frequencies and intensities) as a function of disorder, proposing a microscopic understanding based on bond polarizabilities.

\section{Computational details}
\label{sec:computational-details}
The dynamical properties as well as Raman spectra are calculated with the \textsc{Abinit} software package~\cite{gonze2002first,gonze2009abinit,gonze2016recent}.
Norm-conserving (NC) pseudopotentials are used with Cu(3s$^2$3p$^6$3d$^{10}$4s$^1$), Zn(3s$^{2}$3p$^{6}$3d$^{10}$4s$^2$), Sn(4d$^{10}$5s$^2$5$p^2$) and S(3s$^2$3p$^4$) levels treated as valence states.
These pseudopotentials have been generated using the ONCVPSP package~\cite{hamann2013} and validated within the PseudoDojo framework~\cite{lejaeghere2016reproducibility,van2018pseudodojo}.

In the VCA, the disorder on the different sites of the alloy structure is accounted for by introducing virtual atoms whose pseudopotentials are an \textit{average} of those of the actual constituent atoms in the ordered structure~\cite{van2002effect, zunger1990special}. Several theoretical works~\cite{leung2002effective, serrano2001isotopic, ghosez2000first, mikami2006first, alves2012lattice} have previously successfully employed the VCA to compute the structural, dielectric, piezoelectric, dynamical, and electronic properties of materials, as well as their phase stability. 

Apart from KS, CZ-PD, and CZ-FD phases, we consider a phase CZ-1 that lies between KS and CZ-PD, and two other phases CZ-2 and CZ-3 that lie between CZ-PD and CZ-FD.
Finally, for the sake of completeness, and to better illustrate the trends, we also investigate a series of structures lying between CZ-FD and the stannite (ST) form of CZTS, namely CZ-4, CZ-5, and CZ-6.
We start from KS, and increase the disorder by mixing Cu and Zn within the 2$c$ and 2$d$ sites until CZ-PD, with decreasing (increasing) partial occupancy of Zn (Cu) at the 2$c$ site and the other way around for the 2$d$ site.
Beyond CZ-PD, we also start mixing the Cu atom present at the 2$a$ site with those at the 2$c$ and 2$d$ sites with decreasing (increasing) partial occupation Cu (Zn) in the 2$a$ site leading to CZ-FD.
Extending this disorder further, we ultimately reach the ST phase with Zn occupying the 2$a$ site while the 2$c$ and 2$d$ sites are occupied by Cu.
To conceive the idea of change in the partial occupation within Cu-Zn sites in a general fashion, we use the order parameters $\eta$ and $\xi$ defined in such a way that the occupancy at sites 2$a$, 2$c$, and 2$d$ are given as (Cu$_{1-\eta}$, Zn$_\eta$), (Cu$_\frac{1+\eta-\xi}{2}$, Zn$_\frac{1-\eta+\xi}{2}$), and (Cu$_\frac{1+\eta+\xi}{2}$, Zn$_\frac{1-\eta-\xi}{2}$) respectively. The different values of $\eta$ and $\xi$ for the structures considered here are presented in Table.~\ref{tab:partial-occup}.

\begin{table}
\caption{
Partial occupation (Cu and Zn) of the 2$a$, 2$c$, and 2$d$ sites in CZTS for the different phases considered here, along with the corresponding order parameters ($\eta$ and $\xi$).
}
\centering
\begin{ruledtabular}

\begin{tabular}{l c c c c c c c c c c c c c  }
      & & \multicolumn{2}{c}{2$a$} & & \multicolumn{2}{c}{2$c$} & & \multicolumn{2}{c}{2$d$} \\
      & & Cu   & Zn   & & Cu   & Zn   & &  Cu   & Zn & & $\eta$ & & $\xi$\\
\cline{3-4} \cline{6-7} \cline{9-10} \cline{12-14}
KS    & & 1    & 0    & & 0    & 1    & & 1    & 0    & & 0   & & 1   \\
CZ-1  & & 1    & 0    & & 1/4  & 3/4  & & 3/4  & 1/4  & & 0   & & 1/2 \\
CZ-PD & & 1    & 0    & & 1/2  & 1/2  & & 1/2  & 1/2  & & 0   & & 0   \\
CZ-2  & & 0.90 & 0.10 & & 0.55 & 0.45 & & 0.55 & 0.45 & & 0.1 & & 0   \\
CZ-3  & & 0.80 & 0.20 & & 0.60 & 0.40 & & 0.60 & 0.40 & & 0.2 & & 0   \\
CZ-FD & & 2/3  & 1/3  & & 2/3  & 1/3  & & 2/3  & 1/3  & & 1/3 & & 0   \\
CZ-4  & & 1/2  & 1/2  & & 3/4  & 1/4  & & 3/4  & 1/4  & & 1/2 & & 0   \\
CZ-5  & & 1/4  & 3/4  & & 7/8  & 1/8  & & 7/8  & 1/8  & & 3/4 & & 0   \\
CZ-6  & & 0.10 & 0.90 & & 0.95 & 0.05 & & 0.95 & 0.05  & & 0.9 & & 0   \\
ST    & & 0    & 1    & & 1    & 0    & & 1    & 0    & & 1   & & 0   \\

\end{tabular}

\end{ruledtabular}

\label{tab:partial-occup}
\end{table}

Most of the calculations are performed using the local density approximation (LDA)~\cite{kohn1965self} for evaluating the XC energy. The wavefunctions for the density functional calculations are expanded using a plane-wave basis set up to a kinetic energy cutoff of 50~Ha.
The Brillouin zone is sampled using a 6$\times$6$\times$6 Monkhorst-Pack $k$-point grid~\cite{monkhorst1976special}.
The structures are fully relaxed with a tolerance of 1$\times$10$^{-8}$~Ha/Bohr ($\approx$5$\times$10$^{-4}$~meV/\AA) on the remaining forces.

The phonon frequencies are computed at the $\Gamma$ point of the Brillouin zone.
The Raman scattering efficiency of the phonon of frequency~$\omega_m$ for a photon of frequency~$\omega_i$ is defined as presented in~\citet{veithen2005nonlinear}:
\begin{eqnarray}
I &=& (\omega_i - \omega_m)^4 | \textbf{e}_o . \boldsymbol{\alpha}_m . \textbf{e}_i |^2 \frac{n_m +1}{2 \omega_m}, \label{eq:raman-scattering}
\end{eqnarray}
with $\textbf{e}_i$ and $\textbf{e}_o$ the incoming and outgoing light polarizations, 
$\boldsymbol{\alpha}_m$ the Raman susceptibility, $n_m$ the temperature-dependent phonon occupation factor:
\begin{equation}
n_m = \frac{1}{e^{{\omega_m}/{kT}}-1}.
\end{equation}
The Raman susceptibility $\boldsymbol{\alpha}_m$ is approximated by its static value rather than being evaluated at the frequency of the incoming light and without taking into account the excitonic effects~\cite{Gillet2013}. We take into account the polycrystalline nature of the experimental sample, \ie, considering intensities both in the parallel and perpendicular laser polarizations, and providing an average over all the possible orientations of the crystal as prescribed by~\citet{caracas2009elasticity}. 

For the KS structure, we also compute the phonon frequencies at the $\Gamma$ point using the HSE hybrid XC functional~\cite{heyd2003hybrid,krukau2006influence,payne1992iterative} via finite displacements (or finite differences) as implemented in the \textsc{Phonopy} package~\cite{togo2015first}. The related DFT calculations were performed with the \textsc{Vasp}~\cite{kresse1996efficiency} code using the projector augmented wave (PAW)~\cite{blochl1994projector} method. The atomic positions and lattice vectors were optimized with a convergence criterium on the total energy of 10$^{-8}$~meV/atom and 2~meV/$\mbox{\AA}$ on the forces. We consider a kinetic energy cut-off of 450~eV for truncating the plane-wave basis set and $k$-point density of 750 points per reciprocal atom centered at $\Gamma$. 
Similar to a previous study by some of us~\cite{ramkumar2016first} where the effect of XC functional was assumed not to influence the Raman intensities, the HSE Raman spectrum is determined using the intensities from LDA and NC pseudopotentials computed with \textsc{Abinit}.


\section{Results and discussion}
\label{sec:results-discussion}
\subsection{Phonons of VCA structures}
\label{sec:phonons}

The KS and CZ-1 phases display space group $I\bar{4}$ (N$^\circ$~82). All the other disordered structures are in the space group $I\bar{4}2m$ (N$^\circ$~121), with more symmetries than that of KS, due to the introduction of mirror planes (same occupation of 2$c$ and 2$d$ sites)~\cite{lafond2014x}. 
Group theory analysis representing the acoustic and optic phonon modes at the $\Gamma$ point of Brillouin zone provides the following irreducible representations.
For KS and CZ-1 structures (in the $I\bar{4}$ space group), we have: 
\begin{center}
$\Gamma$=$\underbrace{1B \oplus 1E}_{\text{Acoustic}} \oplus \underbrace{3A \oplus \underbrace{ 6B\oplus6E}_\text{IR}}_{\text{Raman}}$,
\end{center}  
and for the remaining structures (in the $I\bar{4}2m$ space group) it is:
\begin{center}
$\Gamma$=$ \underbrace{1B_{2} \oplus 1E}_{\text{Acoustic}} \oplus \underbrace{2A_{1} \oplus 2B_{1} \oplus \underbrace{ 4B_{2}\oplus6E}_\text{IR}}_{\text{Raman}}  \oplus \underbrace{A_{2}}_{\text{Silent}} $.
\end{center} 
Based on the above, we can identify the Raman active phonon modes for the structures under consideration.
In Table~\ref{tab:phonon-freq}, the computed LDA and HSE phonon frequencies for the KS structure are compared with experimental observations to provide a clear understanding on this aspect.
The LDA TO phonon frequencies of all the structures are provided in Table~\ref{tab:phonon-freq-full} of the Appendix~\ref{app:phonon-freqs}. For the KS structure, the mean absolute difference (MAD) between TO modes of LDA (resp. HSE) with respect to experiment is 14.8~cm$^{-1}$ (resp.~4.1~cm$^{-1}$).
Further, the largest MAD for LDA are from the $A$ modes with 18.9~cm$^{-1}$, while with HSE it is about 10 times smaller with a value 1.8~cm$^{-1}$. This clearly shows the improved prediction of $A$ modes frequencies from HSE,  in agreement with the findings of Ref.~\protect\onlinecite{monserrat2018role}.

Fig.~\ref{fig:raman-spectra} shows the calculated Raman spectra in comparison to the experimental spectrum of~\citet{dimitrievska2014multiwavelength}. 

For frequencies below 150~cm$^{-1}$, both the LDA and HSE spectra are similar to the experimental ones. But above that value, the experimental spectra are well represented only by HSE. 
Therefore we confirm the difference between LDA and HSE spectra due to the peak shifts already observed in Ref.~\protect\onlinecite{monserrat2018role}. However, this does not completely remove the discrepancy with respect to experiments.

Indeed, despite the good agreement of the phonon frequencies using HSE, the Raman intensities of KS structure do not correspond well with the experimental one, especially the prominent $A$ modes ($A$(2) and $A$(3) with respect to KS). A possible reason for this was pointed out in a previous work by some of us~\cite{ramkumar2016first}, where we had ascribed it to the presence of disorder in the system. 

Another possible reason might be erroneous LDA intensities. Hereafter, we will show that the changes in Raman intensity due to disorder, computed within LDA for these particular $A$ modes, are sufficient to understand the characteristics of CZTS Raman spectra.
Given that the structures display different space groups (either $I\bar{4}$ or $I\bar{4}2m$), the three $A$ modes of KS and CZ-1 ($A$(1), $A$(2), and $A$(3)) become $A_2$, $A_1$(1), and $A_1$(2). In order to ease the reading of the rest of the manuscript, hereafter we will refer to these three modes as $A'$, $A''$, and $A'''$.

\begin{table}[h]
\caption{
\label{tab:phonon-freq}
Calculated $\Gamma$-point phonon frequencies (in cm$^{-1}$) for the KS phase of CZTS using LDA and HSE. For the former, both TO and LO frequencies were obtained, while for the latter, only TO frequencies were computed. The experimental Raman (from Refs.~\protect\onlinecite{dimitrievska2014multiwavelength},~\protect\onlinecite{fontane2012vibrational}, and~\protect\onlinecite{guc2016polarized}) and IR (from Ref.~\protect\onlinecite{himmrich1991far}) frequencies are also shown for comparison.
}
\begin{ruledtabular}
\begin{tabular}{l r r r r c r r}
Mode & \multicolumn{2}{c}{Theory} & & \multicolumn{4}{c}{Experiment} \\
& LDA & HSE &
& Ref.~\protect\onlinecite{dimitrievska2014multiwavelength}
& Ref.~\protect\onlinecite{fontane2012vibrational}
& Ref.~\protect\onlinecite{guc2016polarized}
& Ref.~\protect\onlinecite{himmrich1991far}
\\
\cline{2-3} \cline{5-8} \\
$A(1)$    & 301.8 & 273.2 & & 287 & 287 & 276 &     \\
$A(2)$    & 306.5 & 287.8 & & 302 &     & 287 &     \\
$A(3)$    & 326.7 & 339.7 & & 338 & 337 & 338 &     \\
$B(\TO1)$ &\094.3 &\088.4 & &\082 &\083 &\084 &\086 \\
$B(\LO1)$ &\095.4 &       & &     &     &     &     \\
$B(\TO2)$ & 105.9 &\095.5 & &\097 &\097 &     &     \\
$B(\LO2)$ & 106.1 &       & &     &     &     &     \\
$B(\TO3)$ & 178.5 & 170.4 & & 164 & 166 & 167 & 168 \\
$B(\LO3)$ & 178.6 &       & &     &     &     &     \\
$B(\TO4)$ & 268.4 & 233.9 & & 255 & 252 & 248 &     \\
$B(\LO4)$ & 284.6 &       & & 263 &     & 255 &     \\
$B(\TO5)$ & 329.2 & 314.7 & & 331 &     & 311 & 316 \\
$B(\LO5)$ & 333.0 &       & &     &     & 320 &     \\
$B(\TO6)$ & 349.5 & 355.7 & & 353 & 353 & 352 &     \\
$B(\LO6)$ & 359.0 &       & & 374 &     & 374 &     \\
$E(\TO1)$ &\081.8 &\074.5 & &\068 &\066 &\083 &\068 \\
$E(\LO1)$ &\081.8 &       & &     &     &     &     \\
$E(\TO2)$ & 101.8 & 102.1 & &\097 &\097 &\098 &     \\
$E(\LO2)$ & 101.8 &       & &     &     &     &     \\
$E(\TO3)$ & 166.1 & 147.1 & & 140 & 143 & 145 & 143 \\
$E(\LO3)$ & 166.2 &       & & 151 &     &     &     \\
$E(\TO4)$ & 277.3 & 244.4 & &     &     & 250 & 255 \\
$E(\LO4)$ & 288.3 &       & & 271 & 272 & 264 &     \\
$E(\TO5)$ & 312.0 & 286.8 & & 316 &     & 293 & 293 \\
$E(\LO5)$ & 314.4 &       & &     &     & 300 &     \\
$E(\TO6)$ & 331.9 & 346.2 & & 347 & 347 & 347 & 351 \\
$E(\LO6)$ & 342.0 &       & & 366 &     & 366 &     \\
\end{tabular}
\end{ruledtabular}
\end{table}

\subsection{Raman spectra of VCA structures}
\label{sec:raman-spectra}

\begin{figure}[H]
\centering
\includegraphics{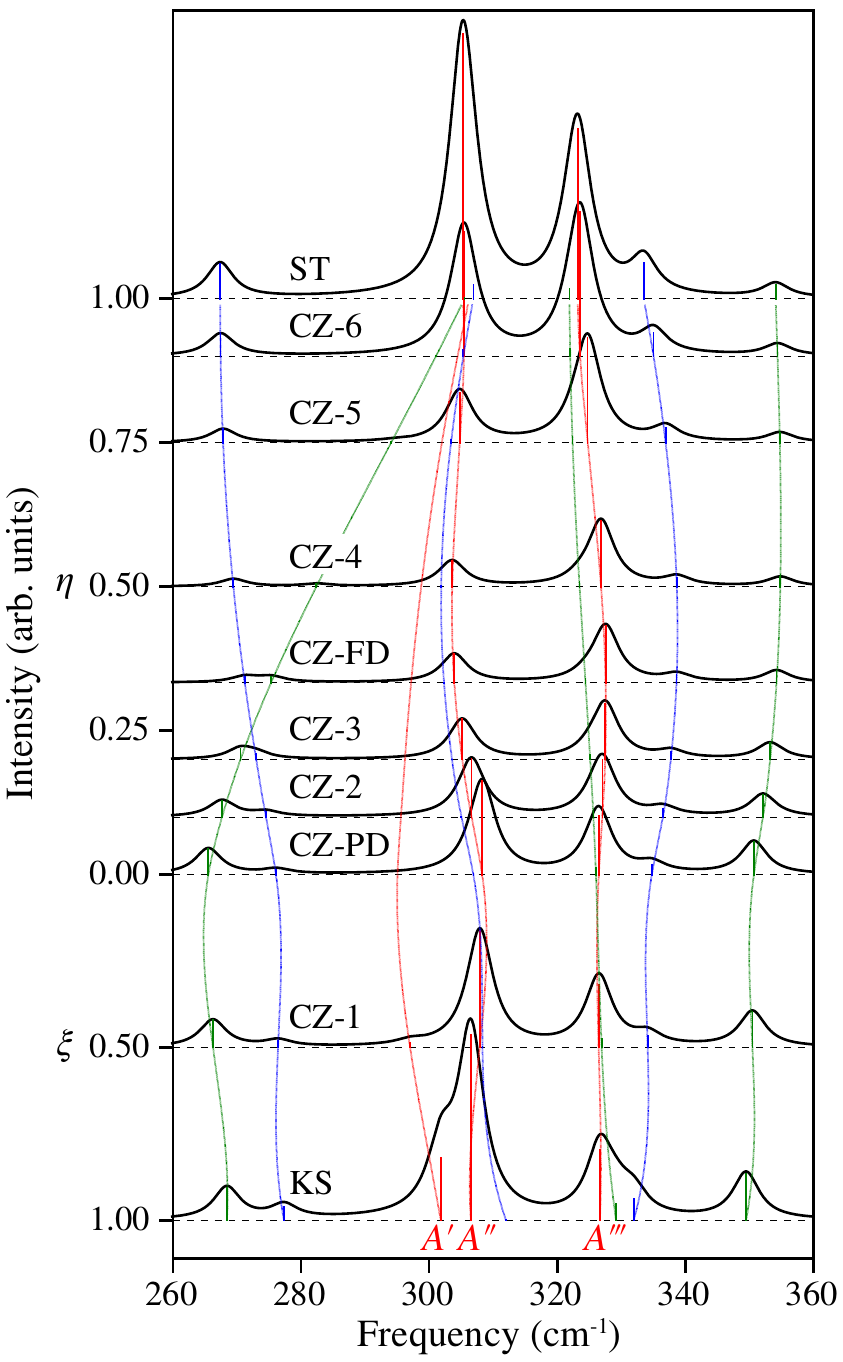}
\caption{CZTS Raman spectra computed using LDA TO modes for KS, CZ-1, CZ-PD, CZ-2, CZ-3, CZ-FD, CZ-4, CZ-5, CZ-6, and ST. 
For comparison with experiment~\cite{dimitrievska2014multiwavelength}, a temperature of 300~K is chosen for the Bose-Einstein occupation factors.
The red, green, and blue vertical lines correspond to the $A$, $B$, and $E$ modes (with the $A'$, $A''$, and $A'''$ ones explicitly indicated at the bottom).
The thinner lines show the evolution of the frequencies from one structure to another.
}
\label{fig:Raman-spectra-normal}
\end{figure}

As mentioned in Section~\ref{sec:phonons}, the discrepancy between theory and experiment in Raman intensities of higher frequency $A$ modes, and its link to the nature of disorder present in CZTS are two important points that need further investigation. 
Fig.~\ref{fig:Raman-spectra-normal} shows the Raman spectra for the structures considered starting from KS as a function of $\eta$ and $\xi$ from 260~cm$^{-1}$ onward. The spectra below 260~cm$^{-1}$ do not show a significant change as can be seen from the log plot of Raman spectra in Fig.~\ref{fig:raman-log} of the Appendix~\ref{app:raman-log}.
The colored lines in Fig.~\ref{fig:Raman-spectra-normal} denote the evolution of different modes where the $B$ (in green) and $E$ (in blue) modes show a larger change in their phonon frequencies with increasing disorder compared to the $A$ modes (red line) which show only a marginal change in their phonon frequencies.
Hence, it cannot reasonably be expected that using HSE, which produces accurate frequencies in comparison to experiment, would lead to a crossing of the frequencies of the two prominent $A$ modes.
 
On the other hand, the intensities of the two prominent $A$ modes ($A''$ and $A'''$) start to change when disorder increases on the Cu-Zn sites.
We can notice this clearly while transitioning from CZ-1 ($\eta=0$) to CZ-FD ($\eta=1/3$).
The intensities of the peaks closer to CZ-FD resemble the experimental ones in Fig.~\ref{fig:raman-spectra} where the lower frequency mode $A''$ has a smaller intensity compared to the higher frequency one $A'''$.
This observation is significant from a theoretical perspective as it further confirms the presence of disorder in all of the Cu-Zn sites as noted by previous works~\cite{bosson2017cation,ramkumar2018insights,wallace2019atomistic,zheng2019cu}.
This solves the discrepancy between earlier theoretical works in comparison to experiments.
Interestingly, with further continuing the mixing process, the system would eventually proceed to the ST phase ($\eta=1$) where the intensities of two $A$ peaks revert back looking similar to that of KS. 

\subsection{Evolution of $A$ modes with disorder}
\label{sec:amodes-evolution}

\begin{figure}[h]
\centering
\includegraphics{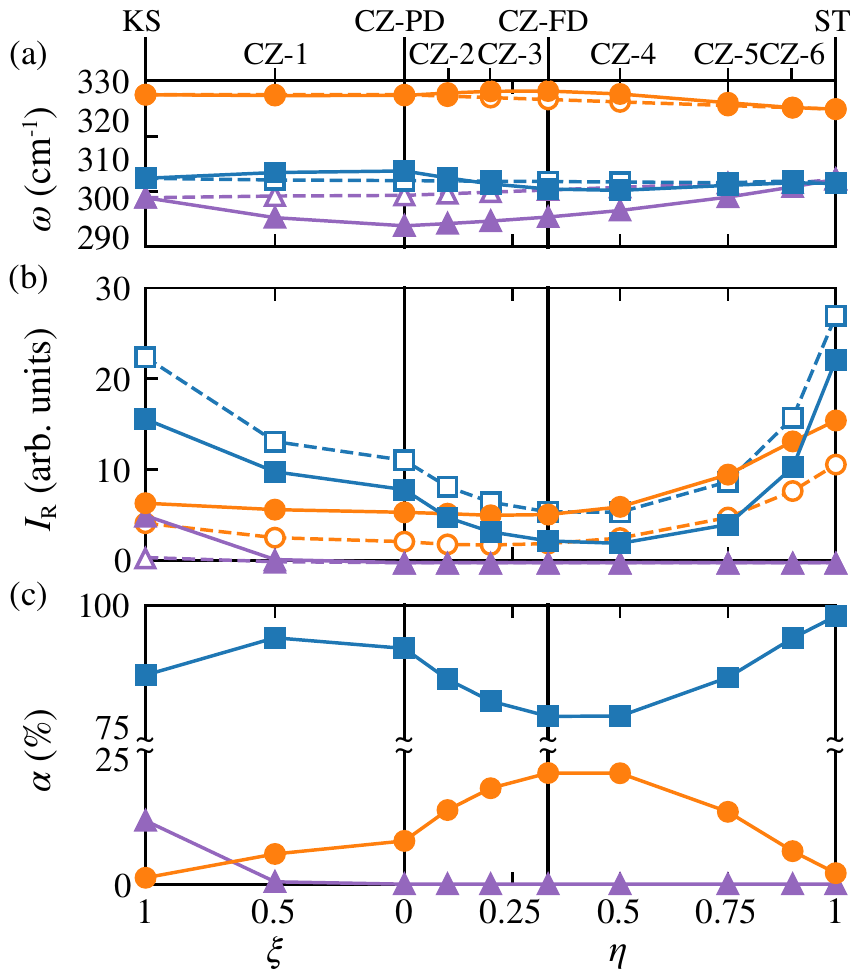}
\caption{Evolution of (a) the phonon frequencies of $A'$ (in purple), $A''$ (in blue), and $A'''$ (in orange) calculated using the VCA (solid lines and markers) and a linear interpolation of the interactomic force constants (dashed lines and open markers), (b) the Raman intensities of $A'$, $A''$, and $A'''$ (solid lines) and $\tilde{A}'$, $\tilde{A}''$, $\tilde{A}'''$ (dashed lines), and (c) the percentage of overlap of $A'$, $A''$, and $A'''$ modes with the ideal $\tilde{A}''$ mode as a function of disorder from KS through the intermediate phases until ST. The interpolation method in (a) deviates from the VCA values indicating its limitation in the prediction of vibrational properties. Also, the Raman intensities of ideal $A$ modes in (b) indicate a possible shortcoming of the BPM model. 
}
\label{fig:Amode-characteristics}
\end{figure}

In order to investigate the origin of the change in the intensity of the two prominent $A$ modes ($A''$ and $A'''$), we consider the evolution of different quantities related to all the $A$ modes along the transition from KS to CZ-PD, then CZ-FD, and finally ST.
First, in Fig.~\ref{fig:Amode-characteristics} (a), we notice that the frequency change of the two prominent $A$ modes ($A''$ blue solid line with filled squares and $A'''$ orange solid line with filled circles) is rather small when going from KS to ST as was also observed in Fig.~\ref{fig:Raman-spectra-normal}.
In fact, these changes in the frequencies are quite well captured by linearly interpolating the interatomic force constants obtained for KS, CZ-PD, and ST (see the corresponding dashed lines with open symbols in the figure and Appendix~\ref{app:avg-property}) instead of relying on explicit calculations within the VCA for intermediate concentrations.
In contrast, for the first mode $A'$ (purple solid line with filled triangles), the frequency changes noticeably by almost 10 cm$^{-1}$, and these changes are not captured by the same interpolation (purple dashed line with open triangles).

In Fig.~\ref{fig:Amode-characteristics}(b), we see that the absolute intensities show a significant change.
The intensity of the $A'$ mode decreases up to CZ-PD and then the mode becomes silent ($A_2$ in the $I\bar{4}2m$ space group) while for the two prominent $A$ modes, the intensities first decrease until CZ-4 (resp. CZ-3) for $A''$ (resp. $A'''$).
The reduction is, however, much more pronounced (by about one order of magnitude) for the $A''$ mode than for the $A'''$ mode, so that the former becomes less intense than the latter just before reaching CZ-2. 
After the minimum, intensities of both modes increase but, once again, the intensity of $A''$ is more pronounced so that it becomes again the more intense only just before ST. 

\begin{figure*}[h]
\centering
\includegraphics{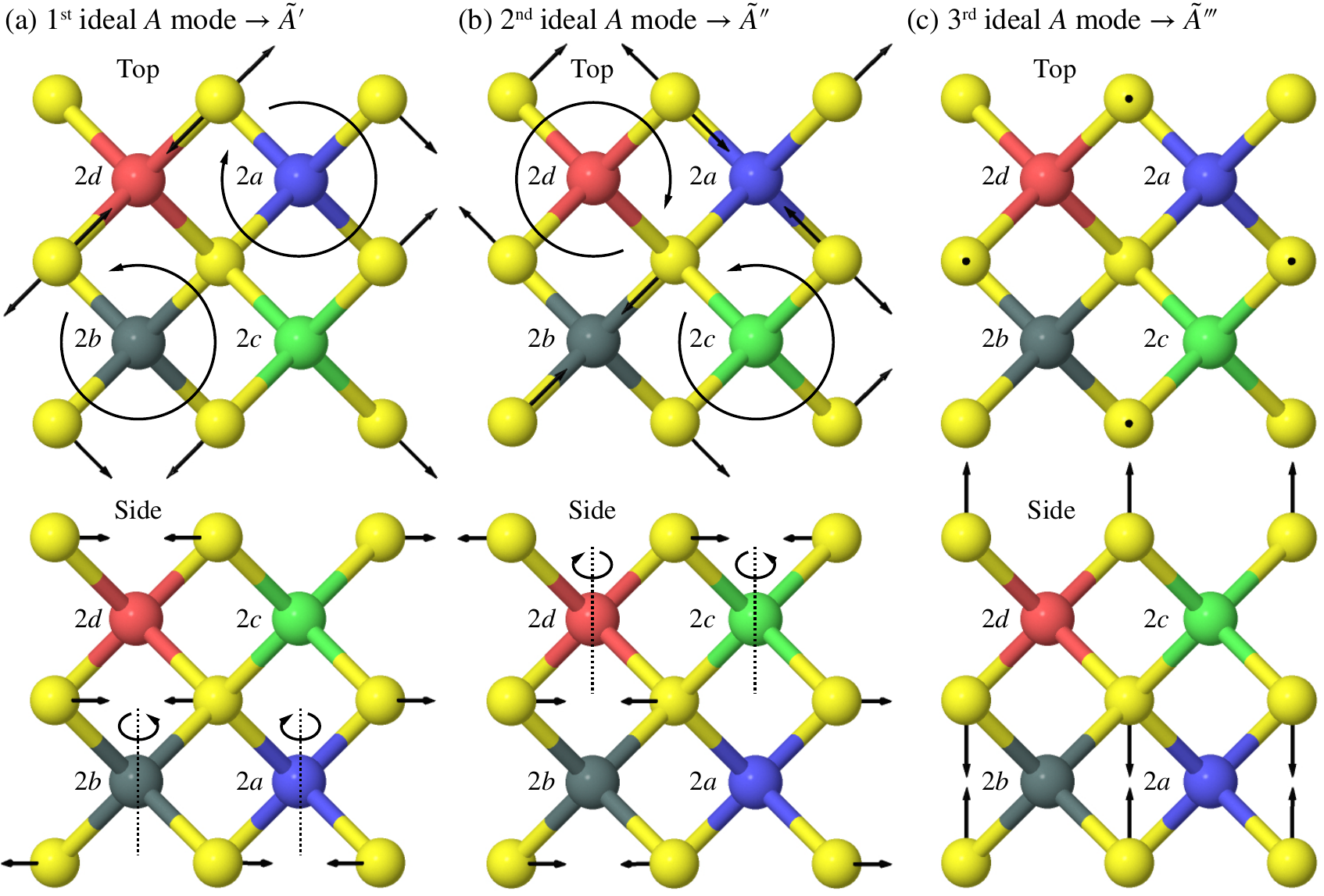}
\caption{Top and side views of the motion of the S atoms for the three ideal $A$ modes ($\tilde{A}'$, $\tilde{A}''$, and $\tilde{A}'''$).
The circular arcs with arrows indicate a rigid rotation of a tetrahedron around an vertical axis reported as a dashed line in the side view.
The Sn and S atoms are in grey and yellow respectively, while the atoms at the 2$a$, 2$c$, and 2$d$ sites are in blue, green, and red.
Their partial occupations are given in Table~\ref{tab:partial-occup}.
}
\label{fig:mode-schematic}
\end{figure*}

Next, we analyze the contribution of the different atomic species to the phonon modes.
A detailed analysis is given in Appendix~\ref{app:atomic-decomp} for KS, CZ-PD, and CZ-FD.
It reveals that the $A$ modes only involve displacements of the S atoms (see Fig.~\ref{fig:atomic_decomp} of the Appendix~\ref{app:atomic-decomp}).
We thus focus on the motion of those atoms and describe them with respect to the tetrahedra that they form with the first-nearest-neighbor cations sitting on the 2$a$, 2$c$, and 2$d$ sites present at the center of tetrahedra.
More specifically, they can be expressed exactly as a linear combination of three ideal $A$ modes (labeled $\tilde{A}'$, $\tilde{A}''$, and $\tilde{A}'''$) as schematically presented in Fig.~\ref{fig:mode-schematic}.
Such a decomposition is particularly interesting for studying the Raman intensities using a bond polarizability model (BPM)~\cite{umari2001raman} where simple considerations can be made.
We note that, for the structures considered in this study, such a model typically involves three bond polarizability parameters (see Appendix~\ref{app:bpm}): the first one is a weighted sum of the longitudinal and perpendicular derivatives of the bond polarizabilities with respect to the bond length (once the former plus twice the latter), the second one is given by their difference, and the third one is related to the difference in the bond polarizabilities.
In the present case, it can be shown that the second one does not contribute to any of the Raman intensities.
Furthermore, the polarizability parameters appear at the second order (either squared or as a mixed product).

Within such a BPM, the rigid rotation of a tetrahedron does not contribute at all to the Raman intensity. As a result, the tetrahedra with the Sn atom and the atom sitting on the 2$a$ site (in grey and blue, respectively) at their center do not contribute to the intensity of the $\tilde{A}'$ mode, which is thus simply related to the difference in the polarizability parameters of the bonds involving the atoms sitting on the 2$c$ and 2$d$ sites (in green and red, respectively) because the S atoms surrounding them move in opposite directions.
From CZ-PD on, these two atoms are exactly the same (and so do the corresponding bond polarizability parameters) and thus the $\tilde{A}$ mode becomes silent.
Similarly, the tetrahedra with the atoms sitting on the 2$c$ and 2$d$ sites at their center do not contribute to the intensity of the $\tilde{A}''$ mode, which depends solely on the difference in the polarizability parameters of the bonds involving the Sn atom and the atom sitting on the 2$a$ site.
Finally, the intensity of the $\tilde{A}'''$ mode will depend on the difference between, on the one hand, the sum of the polarizability parameters of the bonds involving the Sn atom and the atom sitting on the 2$a$ site, and, on the other hand, the sum of the polarizability parameters of the bonds involving the atoms sitting on the 2$c$ and 2$d$ sites.

Fig.~\ref{fig:Amode-characteristics}(c) shows the projection ($\alpha$) of the three real $A$ modes ($A'$ in purple, $A''$ in blue, and $A'''$ in orange) onto the ideal $\tilde{A}''$ mode with increasing disorder.
We first note that, from CZ-PD on, the first value of $A'$ (in purple) is strictly zero and the last two values add up exactly to 100\% (in blue and orange).
This is because from there on, the $A'$ mode perfectly matches the $\tilde{A}'$ one and it thus silent.
Before CZ-PD, the $A'$ mode is basically a mixing of the $\tilde{A}'$ and the $\tilde{A}''$ ones with a very low fraction of the latter (a bit more than 12\% in KS and less than 0.5\% in CZ-1, as illustrated by the purple filled triangles in the figure). 
For $A''$ (resp. $A'''$), beyond CZ-PD, $\alpha$ gradually decreases then increases (resp. increases then decreases).
But all along, $\alpha$ remains larger for $A''$ (in blue) than for $A'''$ (in orange), with a minimum (resp. maximum) when the structure is at CZ-FD.

From the knowledge of Raman tensors for the $A$ modes and the corresponding projections onto the ideal modes, it is possible to determine the intensities of the latter for the different structures.
These are indicated by dashed lines with open symbols (adopting the same color scheme as above) in Fig.~\ref{fig:Amode-characteristics}(b).
We observe that the intensity of the $\tilde{A}''$ mode (blue dashed line and open squares) is by far the largest.
It is actually larger than the one of the real $A''$ mode.
The intensity of first ideal $A$ mode ($\tilde{A}'$) is almost zero everywhere indicating that the polarizability parameters of the bonds involving the atoms (in green and red) sitting on the 2$c$ and 2$d$ sites are not very different even in KS for which they are 100\% Zn and Cu respectively.
Typically, their difference will be smaller than one, so that the second order is even smaller.
Finally, the intensity of the $\tilde{A}'''$ mode is clearly non zero but smaller than the one of real $A'''$ mode.
Furthermore, their differences are smaller for KS than ST.
Given that its intensity is related to the difference between the polarizability parameters of the bonds involving the Sn atom and the atom sitting on the 2$a$ site, this means that the polarizability parameters of the bonds involving Zn atoms are smaller than those involving Cu atoms and, for both of them, the parameters are smaller than those involving Sn atoms.
Note also that the difference between the intensities of the $\tilde{A}''$ and $\tilde{A}'''$ modes is the smallest for CZ-FD that is when the atoms sitting on the 2$a$, 2$c$, and 2$d$ are all the same.
This is also perfectly understandable from the BPM.
It is also to be noted that the Raman intensities calculated from BPM follows the trend obtained from the calculated intensities from DFPT, and hence could be used as a guideline but the inversion of intensities between $A''$ and $A'''$ is not observed indicating a possible shortcoming of the method.

Based on all of the above, the fact that the intensity of the $A''$ mode becomes smaller than the one of the $A'''$ mode can be ascribed to the variation of their projection $\alpha$ on the ideal $\tilde{A}''$.
Indeed, when $\alpha$ increases for the $A'''$ mode (and hence decreases for the $A''$ mode), its intensity increases since the intensity of the $\tilde{A}''$ mode is always bigger than the one of the $\tilde{A}'''$ mode.
At the same time, the intensities of $\tilde{A}''$ and $\tilde{A}'''$ become closer to one another so that the changes in $\alpha$ are sufficient to induce an inversion of order of $A''$ and $A'''$ mode intensities.
Note that, in Appendix~\ref{app:born-charges}, we also provide an analysis of the changes in Raman intensity of $A''$ and $A'''$ with respect to atomic occupation in the different Wyckoff positions in terms of their anomalous Born effective charges. These can indeed be valuable to understand the charge redistribution during mode vibration which can affect the Raman intensity.

Based on the analysis above, it is, in fact, not very surprising that the VCA is a sufficiently good approximation to account for the evolution of Raman intensities in CZTS as a function of the disorder.
It is true that, due to its simplicity, the VCA artificially increases the symmetry of the system.
But, the only consequence of this is that the Raman intensity of the $A'$ goes to zero, luckily not those of the $A''$ and $A'''$ modes which are the most interesting ones.
Furthermore, the intensities of these latter modes are related mostly to the motion of S atoms, and not that of Cu or Zn atoms.
Using averaged Cu-Zn occupations on the $2a$, $2c$, and $2d$ sites most probably has little impact on the motion of S atom.
This explains why the VCA works sufficiently well in CZTS.

\section{Conclusion}
We have analyzed the vibrational characteristics of KS and other disordered phases using VCA. We have observed that the phonon frequencies of the two prominent $A$ modes ($A''$ and $A'''$) do not change significantly with the disorder while the crux of the problem lies in reproducing qualitatively their Raman intensities in comparison to experiments in the literature. Investigating the Raman spectra with increasing disorder in the Cu-Zn sites shows a clear evolution of the $A''$ and $A'''$ Raman peaks from KS to CZ-FD phase agreeing well with the observed experimental spectrum while a continuation of this disordering process ultimately leads to the ST phase. Further characterization of the computed $A$ modes shows that the intensities are directly linked to the bond polarizability parameters within the BPM, the amount to which they represent the $\tilde{A}''$ mode,
and finally their anomalous Born effective charges; all the three aspects linked to each other. The  polarizability parameters for the three modes show how the motion of sulfur atoms contribute based on the different Wyckoff positions, and qualitatively capture the change in Raman intensities. It mainly shows why the $A'$ mode is silent, while the subtler aspects of intensity changes, especially the inversion between $A''$ and $A'''$ were understood by the projection onto $\tilde{A}''$ mode.
Finally, we have proposed a microscopic understanding based on the anomalous Born charges which are linked to the Raman intensity, and shown how the change in Born charges from its nominal value in combination with the polarizability of the atom leads to a change in intensity observed in our calculations.
In summary, by moving beyond the standard low energy structures in the literature, \ie, by considering explicitly Cu-Zn disorder in the system via VCA, we have provided a clear corroboration between experimental and theoretical Raman spectra in CZTS. 
This also shows that the VCA method captures well the changes in Raman intensity with the disorder, and can be employed for characterizing other similar systems. This study hence provides valuable insights to aid further experimental investigations within the CZTS photovoltaic community to analyze the disorder characteristics, as well as indicates the importance of considering theory related to disordering processes in order to obtain a congruence with experimental results.


\begin{acknowledgments}

S.P.R. would like to thank the FRIA grant of the Fonds de la Recherche Scientifique (F.R.S.-FNRS), Belgium; G.-M.R. is grateful to the Walloon Region (DGO6) through the CZTS project of the ''Plan Marshall 2.vert'' program, and F.R.S.-FNRS for the financial support. Computational resources have been provided by the supercomputing facilities of the Universit\'{e} catholique de Louvain (CISM/UCL) and the Consortium des Equipements de Calcul Intensif en F\'{e}d\'{e}ration Wallonie Bruxelles (CECI) funded by the Fonds de la Recherche Scientifique de Belgique (FRS-FNRS). S.P.R would like to thank Elizabeth A. Nowadnick and Anna Miglio for insightful discussions.
\end{acknowledgments}

\appendix

\section{Phonon frequencies at $\Gamma$ of KS, ST and the other disordered structures}
\label{app:phonon-freqs}
\setcounter{figure}{0}
\setcounter{table}{0}
\renewcommand{\thefigure}{A\arabic{figure}}
\renewcommand{\thetable}{A\arabic{table}}

In Table~\ref{tab:phonon-freq-full}, we present the TO phonon frequencies computed using LDA for KS, CZ-1, CZ-PD, CZ-2, CZ-3, CZ-FD, CZ-4, CZ-5, CZ-6, and ST phases where the first two display a $I\bar{4}$ space group, and the rest a $I\bar{4}2m$ space group.
\begin{table*}
\caption{
\label{tab:phonon-freq-full}
Calculated $\Gamma$-point TO phonon frequencies (in cm$^{-1}$) using LDA of CZTS structures namely KS, CZ-1 in the $I\bar{4}$ space group; CZ-PD, CZ-2, CZ-3, CZ-FD, CZ-4, CZ-5, CZ-6, and ST structures $I\bar{4}2m$ space group.
}
\begin{ruledtabular}
\begin{tabular}{l c c l l c c c c c c c c }
\multicolumn{1}{c}{Mode}& \multicolumn{1}{c}{KS}& \multicolumn{1}{c}{CZ-1} & & \multicolumn{1}{c}{Mode} & \multicolumn{1}{c}{CZ-PD} & \multicolumn{1}{c}{CZ-2}& \multicolumn{1}{c}{CZ-3} & \multicolumn{1}{c}{CZ-FD} & \multicolumn{1}{c}{CZ-4} &\multicolumn{1}{c}{CZ-5} &\multicolumn{1}{c}{CZ-6}  &\multicolumn{1}{c}{ST} \\  
\cline{1-3} \cline{5-13} 

 $A(1)$    &  301.8  & 297.0    && $A_2$\2     & 295.1 & 295.6  & 296.2& 297.2  & 298.7  & 302.0  & 304.4  & 306.3  \\
 $A(2)$    &  306.5  & 307.9    && $A_1(1)$    & 308.3 & 306.6  & 305.2& 303.9  & 303.6  & 304.8  & 305.4  & 305.3   \\
 $A(3)$    &  326.7  & 326.5    && $A_1(2)$    & 326.5 & 327.1  & 327.5& 327.6  & 326.8  & 324.7  & 323.6  & 323.2  \\
 $B(\TO1)$ & \094.3  &\094.7    && $B_1(1)$    &\095.0 &\094.6  &\094.1&\093.5  &\092.7  &\091.7  &\091.1  &\090.8   \\
 $B(\TO2)$ &  105.9  & 102.7    && $B_1(2)$    & 326.0 & 100.6  &\099.7& 324.4  & 323.5  & 322.4  & 322.0  & 171.0  \\
 $B(\TO3)$ &  178.5  & 178.3    && $B_2(\TO1)$ & 101.4 & 178.3  & 178.0&\098.8  &\098.0  &\097.2  &\097.0  &\096.9   \\
 $B(\TO4)$ &  268.4  & 266.2    && $B_2(\TO2)$ & 178.2 & 267.7  & 270.5& 177.2  & 175.6  & 173.1  & 171.8  & 171.0  \\
 $B(\TO5)$ &  329.2  & 326.9    && $B_2(\TO3)$ & 265.5 & 325.6  & 325.1& 275.3  & 282.4  & 294.1  & 301.1  & 305.5  \\
 $B(\TO6)$ &  349.5  & 350.5    && $B_2(\TO4)$ & 350.7 & 352.1  & 352.2& 354.2  & 354.8  & 354.8  & 354.4  & 354.1  \\
 $E(\TO1)$ & \081.8  &\081.5    && $E(\TO1)$   &\081.5 &\080.8  &\080.3&\079.6  &\078.9  &\077.8  &\077.1  &\076.5   \\
 $E(\TO2)$ &  101.8  & 102.0    && $E(\TO2)$   & 102.0 & 102.6  & 103.2& 104.0  & 105.0  & 106.6  & 107.7  & 108.5  \\
 $E(\TO3)$ &  166.1  & 166.7    && $E(\TO3)$   & 167.0 & 166.6  & 166.5& 166.6  & 167.3  & 168.7  & 169.5  & 170.1  \\
 $E(\TO4)$ &  277.3  & 276.4    && $E(\TO4)$   & 276.1 & 274.5  & 273.0& 271.2  & 269.4  & 267.8  & 267.4  & 267.4  \\
 $E(\TO5)$ &  312.0  & 308.3    && $E(\TO5)$   & 306.9 & 305.0  & 303.6& 302.3  & 301.9  & 303.4  & 305.3  & 306.9  \\
 $E(\TO6)$ &  331.9  & 334.1    && $E(\TO6)$   & 334.8 & 336.5  & 337.7& 338.6  & 338.7  & 336.9  & 335.0  & 333.5  \\
\end{tabular}
\end{ruledtabular}
\end{table*}

\begin{table*}
\caption{
\label{tab:raman-coeff}
Coefficients ($a$ and $b$) of the Raman tensor and corresponding intensities $I/C$ (where $C$ is Raman prefactor) for the $A$ and $\tilde{A}$ modes calculated using LDA for the different CZTS structures.
All the numbers have been multiplied by $\times$10$^3$.
}
\begin{ruledtabular}
\begin{tabular}{lrrrrrrrrrrrrrrrrrr}
      & \multicolumn{3}{c}{1$^{st}$ $A$ mode}
      & \multicolumn{3}{c}{2$^{nd}$ $A$ mode}
      & \multicolumn{3}{c}{3$^{rd}$ $A$ mode}
      & \multicolumn{3}{c}{1$^{st}$ $\tilde{A}$ mode}
      & \multicolumn{3}{c}{2$^{nd}$ $\tilde{A}$ mode}
      & \multicolumn{3}{c}{3$^{rd}$ $\tilde{A}$ mode}
      \\
      & \multicolumn{1}{c}{$a^2$} & \multicolumn{1}{c}{$b^2$} & \multicolumn{1}{c}{$I/C$}
      & \multicolumn{1}{c}{$a^2$} & \multicolumn{1}{c}{$b^2$} & \multicolumn{1}{c}{$I/C$}
      & \multicolumn{1}{c}{$a^2$} & \multicolumn{1}{c}{$b^2$} & \multicolumn{1}{c}{$I/C$}
      & \multicolumn{1}{c}{$a^2$} & \multicolumn{1}{c}{$b^2$} & \multicolumn{1}{c}{$I/C$}
      & \multicolumn{1}{c}{$a^2$} & \multicolumn{1}{c}{$b^2$} & \multicolumn{1}{c}{$I/C$}
      & \multicolumn{1}{c}{$a^2$} & \multicolumn{1}{c}{$b^2$} & \multicolumn{1}{c}{$I/C$} \\
\cline{2-4}\cline{5-7}\cline{8-10}\cline{11-13}\cline{14-16}\cline{17-19}
KS    &  4.05 &  1.03 & 35.04 & 12.21 &  2.49 &106.80 &  4.87 &  3.33 & 44.24 &  0.45 &  0.16 &  3.85 & 17.59 &  4.09 &152.90 &  3.09 &  2.60 & 29.34 \\
CZ-1  &  0.27 &  0.12 &  2.34 &  7.73 &  1.57 & 67.61 &  4.36 &  2.92 & 39.45 &  0.08 &  0.05 &  0.75 & 10.41 &  2.71 & 90.05 &  1.86 &  1.86 & 18.61 \\
CZ-PD &  0.00 &  0.00 &  0.00 &  6.17 &  1.14 & 54.31 &  4.17 &  2.67 & 37.45 &  0.00 &  0.00 &  0.00 &  8.78 &  2.22 & 76.09 &  1.55 &  1.59 & 15.67 \\
CZ-2  &  0.00 &  0.00 &  0.00 &  3.82 &  0.72 & 33.61 &  3.98 &  2.85 & 36.44 &  0.00 &  0.00 &  0.00 &  6.55 &  2.02 & 56.45 &  1.25 &  1.56 & 13.61 \\
CZ-3  &  0.00 &  0.00 &  0.00 &  2.61 &  0.50 & 22.92 &  3.76 &  2.99 & 35.24 &  0.00 &  0.00 &  0.00 &  5.23 &  1.89 & 45.04 &  1.14 &  1.60 & 13.12 \\
CZ-FD &  0.00 &  0.00 &  0.00 &  1.84 &  0.35 & 16.19 &  3.71 &  3.29 & 35.78 &  0.00 &  0.00 &  0.00 &  4.37 &  1.85 & 37.76 &  1.19 &  1.79 & 14.21 \\
CZ-4  &  0.00 &  0.00 &  0.00 &  1.64 &  0.33 & 14.37 &  4.24 &  3.98 & 41.53 &  0.00 &  0.00 &  0.00 &  4.33 &  2.02 & 37.59 &  1.55 &  2.29 & 18.31 \\
CZ-5  &  0.00 &  0.00 &  0.00 &  3.26 &  0.73 & 28.36 &  6.80 &  6.01 & 65.50 &  0.00 &  0.00 &  0.00 &  6.97 &  2.89 & 60.23 &  3.09 &  3.85 & 33.63 \\
CZ-6  &  0.00 &  0.00 &  0.00 &  8.12 &  1.76 & 70.86 &  9.67 &  7.51 & 90.17 &  0.00 &  0.00 &  0.00 & 12.50 &  3.88 &107.75 &  5.30 &  5.39 & 53.28 \\
ST    &  0.00 &  0.00 &  0.00 & 17.07 &  2.92 &150.98 & 11.80 &  7.40 &105.73 &  0.00 &  0.00 &  0.00 & 21.01 &  4.34 &183.76 &  7.86 &  5.98 & 72.95 \\
\end{tabular}
\end{ruledtabular}
\end{table*}

\section{Raman spectra in the log scale}
\label{app:raman-log}
\setcounter{figure}{0}
\setcounter{table}{0}
\renewcommand{\thefigure}{B\arabic{figure}}
\renewcommand{\thetable}{B\arabic{table}}

\begin{figure}[H]
\centering
\includegraphics{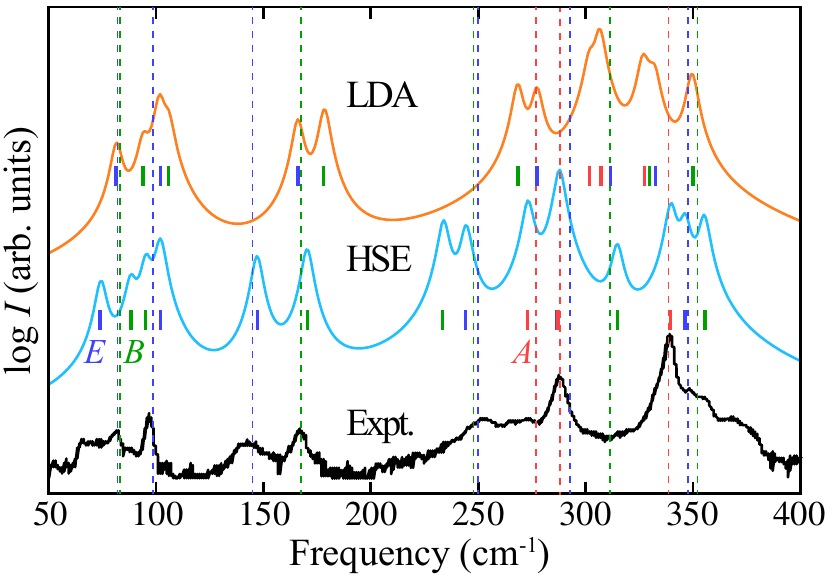}
\caption{Log plot of the Raman spectra computed using LDA and HSE (using LDA intensities) for the KS structure. 
For comparison with experiment~\cite{dimitrievska2014multiwavelength} (black curve in the bottom), a temperature of 300~K is chosen for the Bose-Einstein occupation factors. The red, green and blue dotted lines denote the TO frequencies of A, B, and E modes obtained from polarized Raman measurements made by~\citet{guc2016polarized}.
}
\label{fig:ks-raman-log}
\end{figure}

Fig.~\ref{fig:ks-raman-log} shows the comparison of the calculated Raman spectrum of KS using LDA and HSE (using LDA intensities) to the experimental spectrum of Ref.~\onlinecite{dimitrievska2014multiwavelength} in logarithmic scale. 
While the phonon frequencies are reproduced extremely well with HSE, the calculated intensities do not coincide with experimental observations providing the necessity to consider disorder in the system.

\begin{figure}[!h]
\centering
\includegraphics{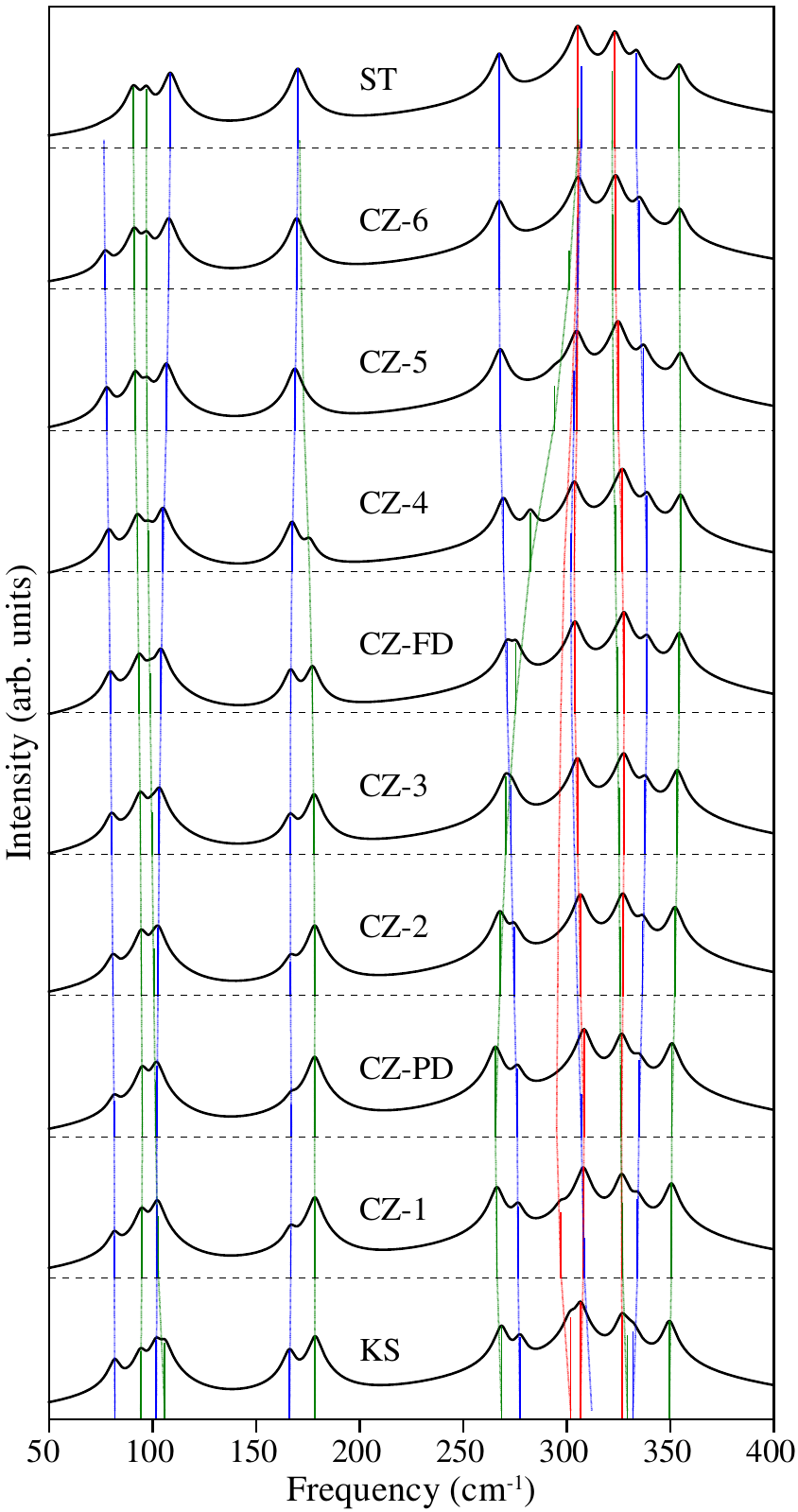}
\caption{Log plot of the Raman spectra as a function of the Cu-Zn disorder starting from KS until ST.
For comparison with experiment~\cite{dimitrievska2014multiwavelength}, a temperature of 300~K is chosen for the Bose-Einstein occupation factors. 
}
\label{fig:raman-log}
\end{figure}

Fig.~\ref{fig:raman-log} shows the logarithmic scale plot of the Raman spectra along with the evolution of $A$ modes (in red), B (in green), and E (in blue) for all the structures considered in this study.
Such plots presented in logarithmic scale yield an easier identification of the low-intensity Raman peaks especially the ones below 200~cm$^{-1}$.

\section{Atomic decomposition of phonon modes in KS, CZ-PD, and CZ-FD}
\label{app:atomic-decomp}
\setcounter{figure}{0}
\setcounter{table}{0}
\renewcommand{\thefigure}{C\arabic{figure}}
\renewcommand{\thetable}{C\arabic{table}}

\begin{figure*}
\includegraphics{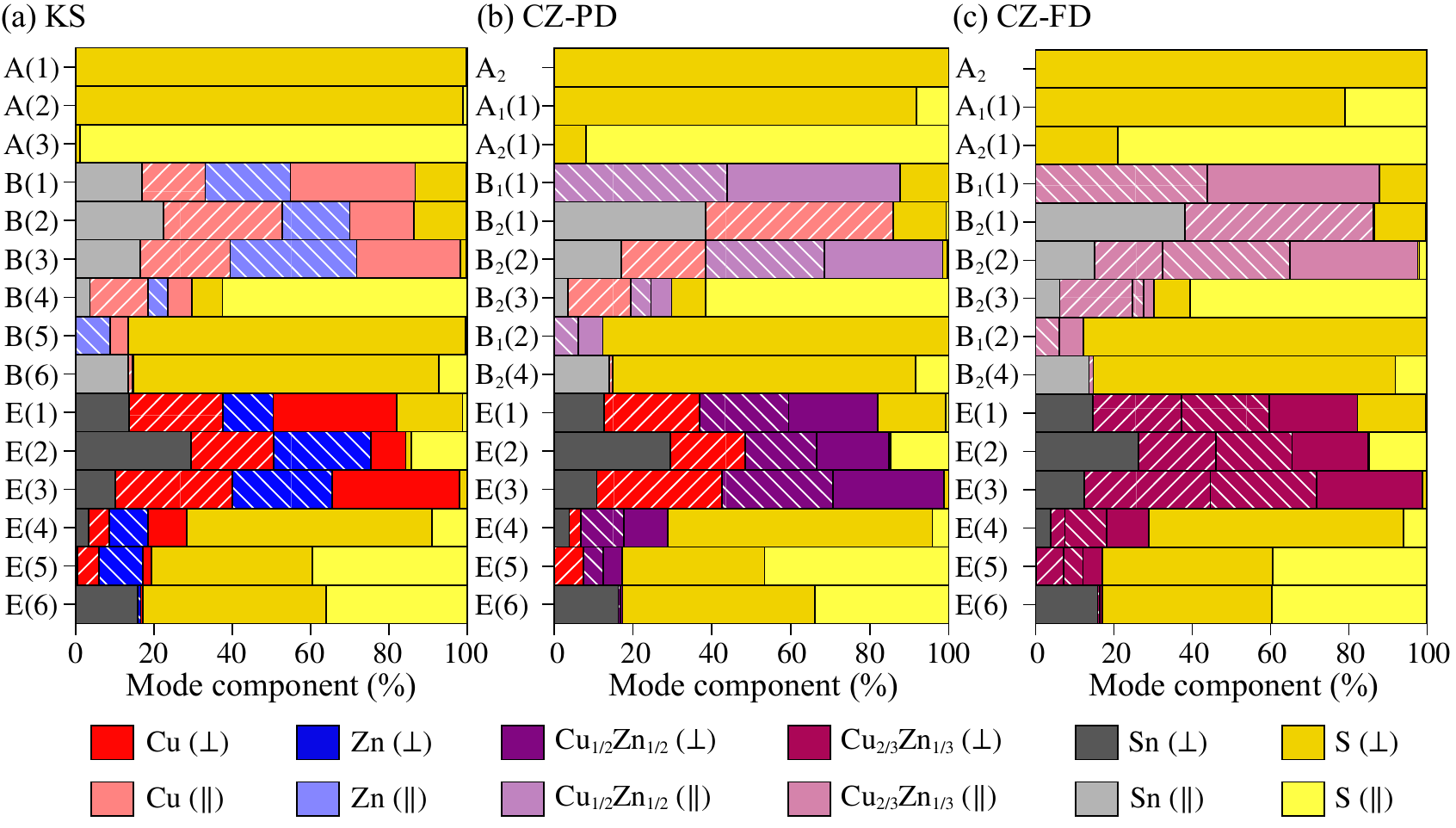}
\caption{Atomic decomposition of the different vibrational modes as computed from LDA. The contribution from the Cu, Zn, Sn, and S atoms are shown in red, blue, grey, and yellow respectively. Furthermore, the Cu-Zn sites are distinguished as forward slashes ('/'), backward slashes ('\textbackslash') and plain colors denote the 2$a$, 2$c$ and 2$d$ sites respectively. In CZ-PD and CZ-FD, the violet and maroon colors are the result of percentage of Cu(red) and Zn(blue) partial occupations.
The parallel ($\parallel$) and perpendicular ($\perp$) components to the $c$ axis are represented using light and dark shades of the colors associated with the different atoms. 
}
\label{fig:atomic_decomp}
\end{figure*}

We now analyze the atomic motion associated with the various modes and highlight the similarities and differences between the three structures for which we exploit the normalization condition of the eigendisplacements $U$ as given in Eq.~(51) of~\citet{gonze1997dynamical}:

\begin{equation}
\label{eq:eig-normalize}
\sum_{\kappa i} M_{\kappa}[U_{m}(\kappa i)]^{*}U_{n}(\kappa i)=\delta_{mn},
\end{equation}

where $M_{\kappa}$ is the mass of the ion $\kappa$, $i$ goes over the three cartesian directions, and $m$ and $n$ denote the phonon modes.
We can identify the contribution of each atom for a given mode as presented in Fig.~\ref{fig:atomic_decomp}, in which we decompose the components as parallel ($\parallel$) and perpendicular ($\perp$) to the $c$ axis of the tetragonal cell. In this figure, they are indicated using light and dark shades of the particular color associated with each atom. Further, the 2$a$, 2$c$, and 2$d$ sites are distinguished by forward slashes ('/'), backward slashes ('\textbackslash'), and plain colors respectively.

In Fig.~\ref{fig:atomic_decomp}, many similarities can be observed between KS, CZ-PD, and CZ-FD - the $A$ modes purely involve anion motion, the $B$/$B_1$/$B_2$ combine cation motion $\parallel$ to the $c$ axis and anion motion (both $\parallel$ and $\perp$), and finally the E modes involve cation motion $\perp$ to the $c$ axis while the anion motion is similar to the B-type modes. 
We notice that the atomic contributions of CZ-PD and CZ-FD look alike in many aspects except for a small difference in the $A$ modes compared to the larger differences between than KS and CZ-PD/CZ-FD. A noticeable difference going from CZ-PD to CZ-FD is a decrease of the $\perp$ motion of S atoms in the $A_1$(1) mode while there is an increase of the same in $A_1$(2) mode. This subtle difference can clearly be seen in the overlap matrix shown in Fig.~\ref{fig:overlap_czpd_czfd} of Appendix~\ref{app:atomic-decomp}. This provides us hints that the motion of S atoms, in particular the $A$ modes ($A'$ and $A''$), could be a key factor in helping identify the nature of the disorder in the system.

To analyze this aspect more clearly, similar to Eq.~\ref{eq:eig-normalize}, by using the eigendisplacements associated with two different structures, we determine an overlap between their modes by means of an overlap matrix which provides us a measure of similarity between them. 
The projection $p_{mn}(S_1,S_2)$ of the eigendisplacements $U_{m}^{S_1}$ of structure $S_1$ onto the eigendisplacements $U_{n}^{S_2}$ of structure $S_2$ is given by:
\begin{equation}
p_{mn}(S_1,S_2)=\sum_{\kappa i} M_{\kappa}[U_{m}^{S_1}(\kappa i)]^{*}U_{n}^{S_2}(\kappa i).
\label{eq:mode-projection}
\end{equation}

\begin{figure}[H]
\includegraphics{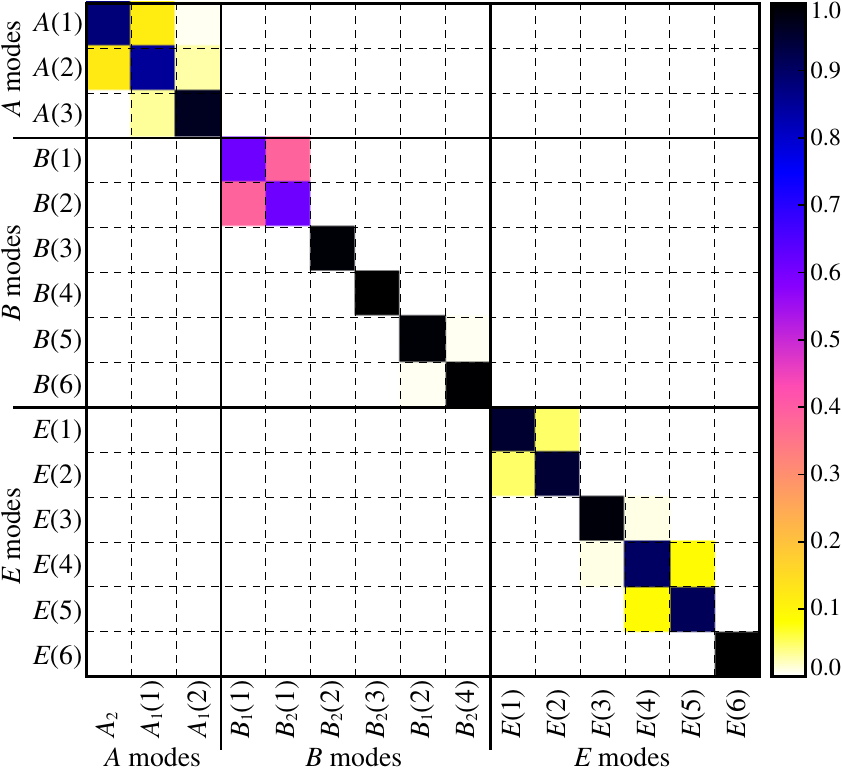}
\caption{Overlap matrix between KS and CZ-PD modes.}
\label{fig:overlap_ks_czpd}
\end{figure}

\begin{figure}[H]
\includegraphics{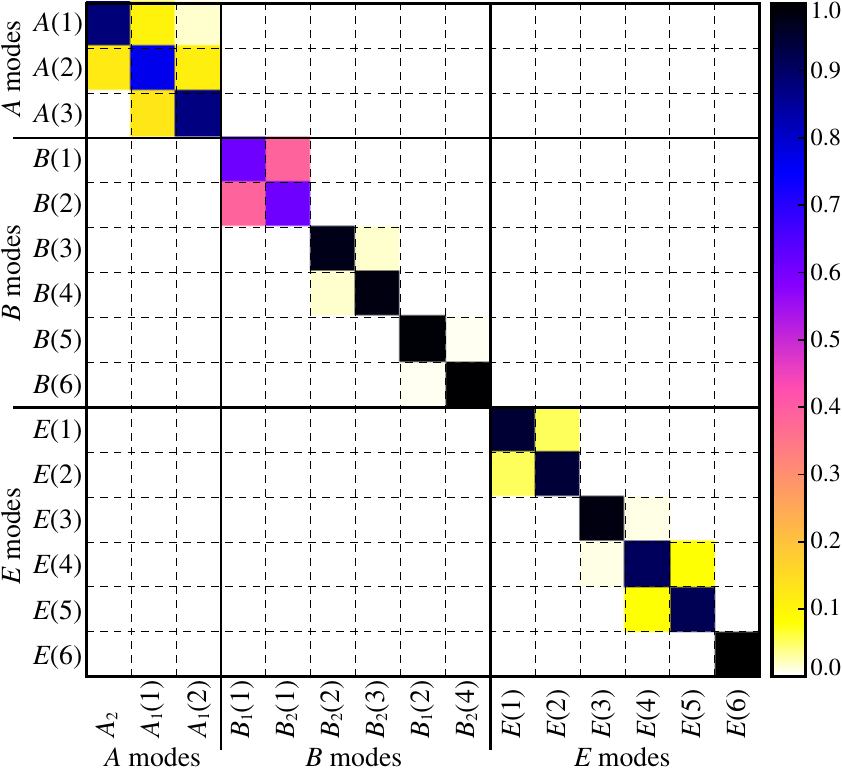}
\caption{Overlap matrix between KS and CZ-FD modes.}
\label{fig:overlap_ks_czfd}
\end{figure}

These overlap matrices between KS - CZ-PD, KS - CZ-FD, and CZ-PD - CZ-FD are shown in Figs.~\ref{fig:overlap_ks_czpd},~\ref{fig:overlap_ks_czfd}, and ~\ref{fig:overlap_czpd_czfd}. The overlap matrix between CZ-PD and CZ-FD shows that they are highly similar except for the slight dissimilarity between their $A$ modes. On the other hand, we find large dissimilarities in the $A$, low-frequency $B$, and a few of the $E$ modes between KS and CZ-PD/CZ-FD.

\begin{figure}[H]
\includegraphics{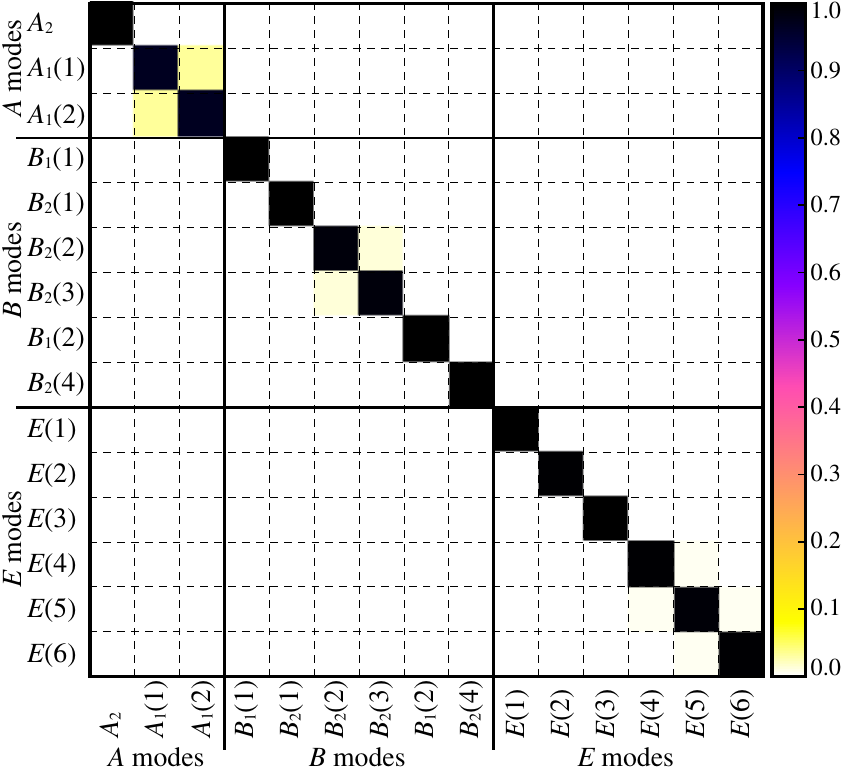}
\caption{Overlap matrix between CZ-PD and CZ-FD modes.}
\label{fig:overlap_czpd_czfd}
\end{figure}

\section{Bond polarizability model for the $A$ modes}
\label{app:bpm}

The Raman susceptibility tensor $\boldsymbol{\alpha}_m$, which appears in Eq.~(\ref{eq:raman-scattering}) is given by:
\begin{equation}
\alpha_{ij}^{m}=\sqrt{\Omega} \sum_{\kappa k} \frac{\partial \chi_{ij}}{\partial \tau_{\kappa k}} U_m(\kappa k)
\label{eq:raman_susc_tensor}
\end{equation}
where $\Omega$ is the volume of the cell, $\tau_{\kappa k}$ are the three cartesian coordinates of atom $\kappa$, and $\chi_{ij}$ is the electric polarizability tensor:
\begin{equation}
\chi_{ij}=\frac{\epsilon^\infty_{ij}-\delta_{ij}}{4\pi}.
\end{equation}

In the bond polarizability model (BPM)~\cite{umari2001raman,tubino1975raman}, it is obtained as the sum of bond contributions:
\begin{equation}
\chi_{ij}=\sum_{\kappa\kappa'} \tilde{\chi}_{ij}(\kappa\kappa').
\end{equation}
And, omitting $(\kappa\kappa')$ in order to simplify the notation, these bond contributions are expressed as:
\begin{equation}
\tilde{\chi}_{ij} =
\frac{2 \beta_p+\beta_l}{3} \delta_{ij} +
\left(\beta_l-\beta_p\right)\left(\frac{R_{i} R_{j}}{R^{2}}-\frac{1}{3} \delta_{i j}\right)
\end{equation}
where $\mathbf{R}=\boldsymbol{\tau}_{\kappa}-\boldsymbol{\tau}_{\kappa'}$ is a vector which defines the direction and the distance of a pair of nearest-neighbor atoms at sites $\kappa$ and $\kappa'$.
The parameters $\beta_l$ and $\beta_p$ correspond to the longitudinal and perpendicular bond polarizability, respectively.

The BPM further assumes that $\beta_l$ and $\beta_p$ only depend on the length of the bond.
Therefore, the derivative of the bond contributions $\tilde{\chi}_{ij}$ with respect to the displacement of the atom $\kappa$ in the direction $k$ is given by:
\begin{eqnarray}
\frac{\partial \tilde{\chi}_{ij}}{\partial \tau_{\kappa k}}&=&
\gamma_1 \delta_{i j} \hat{R}_{k}+
\gamma_2 \left(\hat{R}_i \hat{R}_j-\frac{1}{3} \delta_{ij}\right) \hat{R}_k \nonumber\\
&+&
\gamma_3 \left(\delta_{ik}\hat{R}_j+\delta_{jk}\hat{R}_i-2\hat{R}_i\hat{R}_j\hat{R}_k\right)
\end{eqnarray}
where the three parameters $\gamma_1$, $\gamma_2$, and $\gamma_3$ are defined as:
\begin{equation}
\gamma_1=\frac{2\beta'_p+\beta'_l}{3}, \quad
\gamma_2=\beta'_l-\beta'_p, \quad
\gamma_3=\frac{\beta_l-\beta_p}{R} 
\end{equation}
Each different kind of bond is characterized by three such values.

For the $A$ and $\tilde{A}$ modes, the Raman susceptibility tensors take the following form:
\begin{equation}
\left(
\begin{array}{ccc} a & & \\ & a &  \\ & & b
\end{array}
\right)
\end{equation}
Since only S atoms move in these modes, the only non-zero terms of the sum in Eq.~\ref{eq:raman_susc_tensor} are those for which $\kappa$ is one of these atoms ($U_m(\kappa k)$=0 if $\kappa$ is not one of them).
Hence, in the BPM, the only important contributions are originating from the bonds involving S atoms.
In our structures, there are four types of such bonds: those connecting the S atoms with the atoms sitting on the 2$a$, 2$b$, 2$c$, and 2$d$ sites.
We will thus label the corresponding parameters of the model with a superscript referring to the Wyckoff site.
With this information, it can be shown that:
\begin{widetext}
\begin{align}
&a(\tilde{A}')   = K \left( \Delta_1^{cd} + \Delta_3^{cd} \right) &
&a(\tilde{A}'')  = K \left( \Delta_1^{ab} + \Delta_3^{ab} \right) &
&a(\tilde{A}''') = K \sqrt{2} \left( (\Sigma_1^{ab}-\Sigma_1^{cd}) + 2 (\Sigma_3^{ab}-\Sigma_3^{cd}) \right) \\
&b(\tilde{A}')   = K \left( \Delta_1^{cd} - 2 \Delta_3^{cd} \right) &
&b(\tilde{A}'')  = K \left( \Delta_1^{ab} - 2 \Delta_3^{ab} \right) &
&b(\tilde{A}''') = K \sqrt{2} \left( (\Sigma_1^{ab}-\Sigma_1^{cd}) - 4 (\Sigma_3^{ab}-\Sigma_3^{cd}) \right)
\end{align}
\end{widetext}
with $K=\sqrt{\frac{32}{27 M_S}}$ where $M_S$ is the atomic mass of S, $\Delta_i^{xy}=\gamma_i^{2x}-\gamma_i^{2y}$ and $\Sigma_i^{xy}=\gamma_i^{2x}-\gamma_i^{2y}$ for $i$=1, 2, 3 and $x, y$=$a, b, c$.

\section{Average property of disordered structures}
\label{app:avg-property}
\setcounter{figure}{0}
\setcounter{table}{0}
\renewcommand{\thefigure}{E\arabic{figure}}
\renewcommand{\thetable}{E\arabic{table}}

Here in Table~\ref{tab:avg-property} we present the fraction of phases to be combined in order to obtain the property of the resultant disordered phase. We consider a combination of KS and KS$^*$ (obtained by interchanging Cu and Zn in the 2$c$ and 2$d$ sites in KS) to obtain CZ-1 and CZ-PD, while for the rest of the phases a combination of CZ-PD and ST is considered. In this paper, we compute the phonon frequencies (in Fig.~\ref{fig:Amode-characteristics}(a)) of the resultant phase using the fraction on the interatomic force constants and masses of atoms in the corresponding site for phase-1 and phase-2, 
which can roughly be understood as, $\omega_{net} = x \omega_{phase-1} + (1-x) \omega_{phase-2}$, where $x$ denotes the fraction of phase-1, and $\omega$ being the phonon frequency of the respective phase.

\begin{table}
\label{tab:avg-property}
\caption{
The amount of fraction of phases (phase-1 and phase-2) to be combined in order to obtain the desired property of the disordered phase. The KS$^*$ here is obtained by interchanging the Cu and Zn in the 2$c$ and 2$d$ sites in KS.
}
\centering
\begin{ruledtabular}
\begin{tabular}{l c c c c  }
Phase     & \% & Phase-1 & \% & Phase-2\\
\hline
CZ-1  & 0.25   & KS    & 0.75 & KS$^*$ \\
CZ-PD & 0.50   & KS    & 0.50 & KS$^*$ \\ 
CZ-2  & 0.90   & CZ-PD & 0.10 & ST \\
CZ-3  & 0.80   & CZ-PD & 0.20 & ST \\
CZ-FD & 2/3    & CZ-PD & 1/3  & ST \\
CZ-4  & 0.50   & CZ-PD & 0.50 & ST \\
CZ-5  & 0.25   & CZ-PD & 0.75 & ST \\
CZ-6  & 0.10   & CZ-PD & 0.90 & ST \\

\end{tabular}
\end{ruledtabular}
\end{table}

\section{Born effective charges}
\label{app:born-charges}
\setcounter{figure}{0}
\setcounter{table}{0}
\renewcommand{\thefigure}{F\arabic{figure}}
\renewcommand{\thetable}{F\arabic{table}}

The Born effective charge (BEC) tensors of the Cu, Zn, Sn, and S atoms for the different structures considered here are presented in Table.~\ref{tab:eff-charges}. BEC describes the linear relation between the induced polarization of the solid along the direction $j$ and the displacement of that atom in the direction $i$, under the condition of zero electric field~\cite{gonze1997dynamical}.
Its component $Z^*_{ij}$ (with $i, j$=$1,2,3$) can also be defined as the proportionality coefficient relating, at linear order, the force on that atom in the direction $i$ due and the homogeneous effective electric field along the direction $j$.

In the 2$b$ site, the nominal charge of the Sn atom does not change much with the disorder which is obvious because the disorder considered here is within the Cu-Zn sites. We notice a small anisotropy in KS and CZ-1. The off-diagonal elements arise when there is a deviation from the tetrahedral symmetry. In the 2$a$ site, the BEC values almost continuously increase with disorder except for a slight decrease in CZ-1 and CZ-PD. Initially with a large anisotropy in KS, it dies out exponentially with disorder indicating the cations in the 2$a$ sites are stabilized in the tetrahedral sites. The diagonal values start close to the nominal charge Cu for KS, and, with disorder, moves toward Zn for ST as expected based on the partial occupancies shown in Table~\ref{tab:partial-occup}. For the 2$c$ site, the anisotropy decreases until CZ-3 with some fluctuations, increases until CZ-4, then decreases drastically at CZ-5 until strongly increasing going towards ST. The BEC can be seen close to the nominal charge of Zn for KS, and then moves towards Cu for the ST phase. For the 2$d$ site, the BEC starts close to the nominal charge of Cu for KS, changes with disorder due to the mixing of Zn atoms, and finally falls back close to the Cu value. For the 8$g$ sites, the principal values decrease starting with KS, but increases continuously from CZ-PD onwards but always stay relatively close to the nominal charge of S atom. The BEC values in 2$a$, 2$c$, and 2$d$ are almost the same for CZ-FD indicating the equal distribution of charges for the fully disordered Cu-Zn phase. Also, the off-diagonal values for the cationic sites seem to be the lowest for CZ-3 and CZ-FD indicating the stability of atoms in perfect tetrahedral symmetry.

\begin{table*}
\caption{
\label{tab:eff-charges}
Born effective charge tensors of the atoms occupying the Wyckoff positions 2$b$, 2$a$, 2$b$, 2$c$, and 8$g$ for the different structures.
The corresponding principal values are also reported as a vector between brackets, as well as their average value.
Note that the 2$b$ and 8$g$ are always occupied by Sn and S atoms, respectively.
}
\setlength\tabcolsep{0.0pt}
\begin{ruledtabular}
\begin{tabular}{l@{}c@{\hspace{-1.8mm}}c@{\hspace{-1.8mm}}c@{\hspace{-1.8mm}}c@{\hspace{-1.8mm}}c}

 & 2$b$ & 2$a$ & 2$c$ & 2$d$ & 8$g$\\
\hline \\[-6pt]
KS    & $\scriptsize{\begin{array}[t]{c} \left(\begin{array}{rrr} +3.15 &+0.08 &\ms\+0.00 \\ -0.08  & +3.15 &\ms\+0.00 \\ \ms\+0.00 & \ms\+0.00  & +3.24 \end{array}\right)\vspace{4pt}\\ 
      \left[ \begin{array}{rrr} +3.15 &+3.15 &+3.24  \end{array}\right] \vspace{2pt}\\
      +3.18 \end{array}}$
      & $\scriptsize{\begin{array}[t]{c} \left(\begin{array}{rrr} +0.64 &-0.90 &\ms\+0.00 \\ +0.90  & +0.64 &\ms\+0.00 \\ \ms\+0.00 & \ms\+0.00  & +0.82 \end{array}\right)\vspace{4pt}\\
      \left[ \begin{array}{rrr} +0.64 &+0.64 &+0.82  \end{array}\right] \vspace{2pt}\\
      +0.70 \end{array}}$
      & $\scriptsize{\begin{array}[t]{c} \left(\begin{array}{rrr} +0.83 &+0.42 &\ms\+0.00 \\ -0.42  & +0.83 &\ms\+0.00 \\ \ms\+0.00 & \ms\+0.00  & +0.60 \end{array}\right)\vspace{4pt}\\
      \left[ \begin{array}{rrr} +0.83 &+0.83 &+0.60  \end{array}\right] \vspace{2pt}\\
      +0.75 \end{array}}$
      & $\scriptsize{\begin{array}[t]{c} \left(\begin{array}{rrr} +1.98 &+0.35 &\ms\+0.00 \\ -0.35  & +1.98 &\ms\+0.00 \\ \ms\+0.00 & \ms\+0.00  & +2.13 \end{array}\right)\vspace{4pt}\\
      \left[ \begin{array}{rrr} +1.98 &+1.98 &+2.13  \end{array}\right] \vspace{2pt}\\
      +2.03 \end{array}}$
      & $\scriptsize{\begin{array}[t]{c} \left(\begin{array}{rrr} -1.85 &-0.43 &    +0.75 \\ -0.05  & -1.45 &    -0.09 \\     +0.86 &     +0.15  & -1.70 \end{array}\right)\vspace{4pt}\\
      \left[ \begin{array}{rrr} -2.62 &-1.45 &-0.93  \end{array}\right] \vspace{2pt}\\
      -1.67 \end{array}}\vspace{6pt}$
      \\
CZ-1  & $\scriptsize{\begin{array}[t]{c}\left(\begin{array}{rrr} +3.19 &+0.04 & \ms\+0.00 \\-0.04 & +3.19 & \ms\+0.00 \\ \ms\+0.00 & \ms\+0.00 & +3.16 \end{array}\right)\vspace{4pt}\\
      \left[ \begin{array}{rrr} +3.19 &+3.19 &+3.16  \end{array}\right] \vspace{2pt}\\
      +3.18 \end{array}}$
      & $\scriptsize{\begin{array}[t]{c}\left(\begin{array}{rrr} +0.62 &-0.04 & \ms\+0.00 \\+0.04 & +0.62 & \ms\+0.00 \\ \ms\+0.00 & \ms\+0.00 & +0.79 \end{array}\right)\vspace{4pt}\\
      \left[ \begin{array}{rrr} +0.62 &+0.62 &+0.79  \end{array}\right] \vspace{2pt}\\
      +0.68 \end{array}}$
      & $\scriptsize{\begin{array}[t]{c}\left(\begin{array}{rrr} +1.12 &+0.11 & \ms\+0.00 \\-0.11 & +1.12 & \ms\+0.00 \\ \ms\+0.00 & \ms\+0.00 & +1.08 \end{array}\right)\vspace{4pt}\\
      \left[ \begin{array}{rrr} +1.12 &+1.12 &+1.08  \end{array}\right] \vspace{2pt}\\
      +1.11 \end{array}}$
      & $\scriptsize{\begin{array}[t]{c}\left(\begin{array}{rrr} +1.71 &+0.23 & \ms\+0.00 \\-0.23 & +1.71 & \ms\+0.00 \\ \ms\+0.00 & \ms\+0.00 & +1.84 \end{array}\right)\vspace{4pt}\\
      \left[ \begin{array}{rrr} +1.71 &+1.71 &+1.84  \end{array}\right] \vspace{2pt}\\
      +1.75 \end{array}}$
      & $\scriptsize{\begin{array}[t]{c}\left(\begin{array}{rrr} -1.57 &-0.11 &     -0.10 \\-0.28 & -1.75 &     -0.50 \\     -0.35 &     -0.68 & -1.72 \end{array}\right)\vspace{4pt}\\
      \left[ \begin{array}{rrr} -1.47 &-1.14 &-2.43  \end{array}\right] \vspace{2pt}\\
      -1.68 \end{array}}\vspace{6pt}$
      \\
CZ-PD & $\scriptsize{\begin{array}[t]{c}\left(\begin{array}{rrr}+3.20 & \ms\+0.00 & \ms\+0.00 \\ \ms\+0.00 &+3.20 & \ms\+0.00 \\ \ms\+0.00 & \ms\+0.00 &+3.13 \end{array}\right)\vspace{4pt}\\
      \left[ \begin{array}{rrr} +3.20 &+3.20 &+3.13  \end{array}\right] \vspace{2pt}\\
      +3.18 \end{array}}$
      & $\scriptsize{\begin{array}[t]{c}\left(\begin{array}{rrr}+0.61 & \ms\+0.00 & \ms\+0.00 \\ \ms\+0.00 &+0.61 & \ms\+0.00 \\ \ms\+0.00 & \ms\+0.00 &+0.79 \end{array}\right)\vspace{4pt}\\
      \left[ \begin{array}{rrr} +0.61 &+0.61 &+0.79  \end{array}\right] \vspace{2pt}\\
      +0.67 \end{array}}$
      & $\scriptsize{\begin{array}[t]{c}\left(\begin{array}{rrr}+1.42 &     +0.09 & \ms\+0.00 \\     -0.09 &+1.42 & \ms\+0.00 \\ \ms\+0.00 & \ms\+0.00 &+1.49 \end{array}\right)\vspace{4pt}\\
      \left[ \begin{array}{rrr} +1.42 &+1.42 &+1.49  \end{array}\right] \vspace{2pt}\\
      +1.44 \end{array}}$
      & $\scriptsize{\begin{array}[t]{c}\left(\begin{array}{rrr}+1.42 &     +0.09 & \ms\+0.00 \\     -0.09 &+1.42 & \ms\+0.00 \\ \ms\+0.00 & \ms\+0.00 &+1.49 \end{array}\right)\vspace{4pt}\\
      \left[ \begin{array}{rrr} +1.42 &+1.42 &+1.49  \end{array}\right] \vspace{2pt}\\
      +1.44 \end{array}}$
      & $\scriptsize{\begin{array}[t]{c}\left(\begin{array}{rrr}-1.66 &     -0.18 &     +0.29 \\     -0.18 &-1.66 &     +0.29 \\     +0.52 &     +0.52 &-1.73 \end{array}\right)\vspace{4pt}\\
      \left[ \begin{array}{rrr} -1.49 &-2.37 &-1.20  \end{array}\right] \vspace{2pt}\\
      -1.69 \end{array}}\vspace{6pt}$
      \\
CZ-2  & $\scriptsize{\begin{array}[t]{c}\left(\begin{array}{rrr}+3.18 &\ms\+0.00 &\ms\+0.00 \\ \ms\+0.00 &+3.18 &\ms\+0.00 \\ \ms\+0.00 & \ms\+0.00 & +3.16 \end{array}\right)\vspace{4pt}\\
      \left[ \begin{array}{rrr} +3.18 &+3.18 &+3.16  \end{array}\right] \vspace{2pt}\\
      +3.17 \end{array}}$
      & $\scriptsize{\begin{array}[t]{c}\left(\begin{array}{rrr}+0.81 &\ms\+0.00 &\ms\+0.00 \\ \ms\+0.00 &+0.81 &\ms\+0.00 \\ \ms\+0.00 & \ms\+0.00 & +0.90 \end{array}\right)\vspace{4pt}\\
      \left[ \begin{array}{rrr} +0.81 &+0.81 &+0.90  \end{array}\right] \vspace{2pt}\\
      +0.84 \end{array}}$
      & $\scriptsize{\begin{array}[t]{c}\left(\begin{array}{rrr}+1.36 &    +0.05 &\ms\+0.00 \\     -0.05 &+1.36 &\ms\+0.00 \\ \ms\+0.00 & \ms\+0.00 & +1.40 \end{array}\right)\vspace{4pt}\\
      \left[ \begin{array}{rrr} +1.36 &+1.36 &+1.40  \end{array}\right] \vspace{2pt}\\
      +1.37 \end{array}}$
      & $\scriptsize{\begin{array}[t]{c}\left(\begin{array}{rrr}+1.36 &    +0.05 &\ms\+0.00 \\     -0.05 &+1.36 &\ms\+0.00 \\ \ms\+0.00 & \ms\+0.00 & +1.40 \end{array}\right)\vspace{4pt}\\
      \left[ \begin{array}{rrr} +1.36 &+1.36 &+1.40  \end{array}\right] \vspace{2pt}\\
      +1.37 \end{array}}$
      & $\scriptsize{\begin{array}[t]{c}\left(\begin{array}{rrr}-1.68 &    -0.23 &    -0.28 \\     -0.23 &-1.68 &    -0.28 \\     -0.43 &     -0.43 & -1.72 \end{array}\right)\vspace{4pt}\\
      \left[ \begin{array}{rrr} -1.44 &-2.32 &-1.30  \end{array}\right] \vspace{2pt}\\
      -1.69 \end{array}}\vspace{6pt}$
      \\
CZ-3  & $\scriptsize{\begin{array}[t]{c}\left(\begin{array}{rrr}+3.17 & \ms\+0.00 & \ms\+0.00 \\ \ms\+0.00 & +3.17 & \ms\+0.00 \\ \ms\+0.00 & \ms\+0.00 & +3.18 \end{array}\right)\vspace{4pt}\\
      \left[ \begin{array}{rrr} +3.17 &+3.17 &+3.18  \end{array}\right] \vspace{2pt}\\
      +3.17 \end{array}}$
      & $\scriptsize{\begin{array}[t]{c}\left(\begin{array}{rrr}+0.99 & \ms\+0.00 & \ms\+0.00 \\ \ms\+0.00 & +0.99 & \ms\+0.00 \\ \ms\+0.00 & \ms\+0.00 & +1.02 \end{array}\right)\vspace{4pt}\\
      \left[ \begin{array}{rrr} +0.99 &+0.99 &+1.02  \end{array}\right] \vspace{2pt}\\
      +1.00 \end{array}}$
      & $\scriptsize{\begin{array}[t]{c}\left(\begin{array}{rrr}+1.30 &     +0.02 & \ms\+0.00 \\     -0.02 & +1.30 & \ms\+0.00 \\ \ms\+0.00 & \ms\+0.00 & +1.31 \end{array}\right)\vspace{4pt}\\
      \left[ \begin{array}{rrr} +1.30 &+1.30 &+1.31  \end{array}\right] \vspace{2pt}\\
      +1.30 \end{array}}$
      & $\scriptsize{\begin{array}[t]{c}\left(\begin{array}{rrr}+1.30 &     +0.02 & \ms\+0.00 \\     -0.02 & +1.30 & \ms\+0.00 \\ \ms\+0.00 & \ms\+0.00 & +1.31 \end{array}\right)\vspace{4pt}\\
      \left[ \begin{array}{rrr} +1.30 &+1.30 &+1.31  \end{array}\right] \vspace{2pt}\\
      +1.30 \end{array}}$
      & $\scriptsize{\begin{array}[t]{c}\left(\begin{array}{rrr}-1.69 &     -0.29 & \ms\+0.00 \\     -0.02 & -1.69 &     -0.26 \\     -0.35 &     -0.35 & -1.70 \end{array}\right)\vspace{4pt}\\
      \left[ \begin{array}{rrr} -1.40 &-2.29 &-1.39  \end{array}\right] \vspace{2pt}\\
      -1.69 \end{array}}\vspace{6pt}$
      \\
CZ-FD & $\scriptsize{\begin{array}[t]{c}\left(\begin{array}{rrr} +3.15 & \ms\+0.00 & \ms\+0.00 \\ \ms\+0.00 & +3.15 & \ms\+0.00 \\ \ms\+0.00 & \ms\+0.00 & +3.21 \end{array}\right)\vspace{4pt}\\
      \left[ \begin{array}{rrr} +3.15 &+3.15 &+3.21  \end{array}\right] \vspace{2pt}\\
      +3.17 \end{array}}$
      & $\scriptsize{\begin{array}[t]{c}\left(\begin{array}{rrr} +1.23 & \ms\+0.00 & \ms\+0.00 \\ \ms\+0.00 & +1.23 & \ms\+0.00 \\ \ms\+0.00 & \ms\+0.00 & +1.16 \end{array}\right)\vspace{4pt}\\
      \left[ \begin{array}{rrr} +1.23 &+1.23 &+1.16  \end{array}\right] \vspace{2pt}\\
      +1.21 \end{array}}$
      & $\scriptsize{\begin{array}[t]{c}\left(\begin{array}{rrr} +1.22 &     +0.04 & \ms\+0.00 \\     -0.04 & +1.22 & \ms\+0.00 \\ \ms\+0.00 & \ms\+0.00 & +1.18 \end{array}\right)\vspace{4pt}\\
      \left[ \begin{array}{rrr} +1.22 &+1.22 &+1.18  \end{array}\right] \vspace{2pt}\\
      +1.21 \end{array}}$
      & $\scriptsize{\begin{array}[t]{c}\left(\begin{array}{rrr} +1.22 &     +0.04 & \ms\+0.00 \\     -0.04 & +1.22 & \ms\+0.00 \\ \ms\+0.00 & \ms\+0.00 & +1.18 \end{array}\right)\vspace{4pt}\\
      \left[ \begin{array}{rrr} +1.22 &+1.22 &+1.18  \end{array}\right] \vspace{2pt}\\
      +1.21 \end{array}}$
      & $\scriptsize{\begin{array}[t]{c}\left(\begin{array}{rrr} -1.71 &     -0.37 &     +0.25 \\     -0.37 & -1.71 &     +0.25 \\     +0.25 &     +0.25 & -1.68 \end{array}\right)\vspace{4pt}\\
      \left[ \begin{array}{rrr} -1.40 &-2.29 &-1.39  \end{array}\right] \vspace{2pt}\\
      -1.70 \end{array}}\vspace{6pt}$
      \\
CZ-4  & $\scriptsize{\begin{array}[t]{c}\left(\begin{array}{rrr} +3.13 & \ms\+0.00 & \ms\+0.00 \\ \ms\+0.00 & +3.13 & \ms\+0.00 \\ \ms\+0.00 & \ms\+0.00 & +3.22  \end{array}\right)\vspace{4pt}\\
      \left[ \begin{array}{rrr} +3.13 &+3.13 &+3.22  \end{array}\right] \vspace{2pt}\\
      +3.16 \end{array}}$
      & $\scriptsize{\begin{array}[t]{c}\left(\begin{array}{rrr} +1.51 & \ms\+0.00 & \ms\+0.00 \\ \ms\+0.00 & +1.51 & \ms\+0.00 \\ \ms\+0.00 & \ms\+0.00 & +1.33 \end{array}\right)\vspace{4pt}\\
      \left[ \begin{array}{rrr} +1.51 &+1.51 &+1.33  \end{array}\right] \vspace{2pt}\\
      +1.45 \end{array}}$
      & $\scriptsize{\begin{array}[t]{c}\left(\begin{array}{rrr} +1.13 &     +0.12 & \ms\+0.00 \\     -0.12 & +1.13 & \ms\+0.00 \\ \ms\+0.00 & \ms\+0.00 & +1.02 \end{array}\right)\vspace{4pt}\\
      \left[ \begin{array}{rrr} +1.13 &+1.13 &+1.02  \end{array}\right] \vspace{2pt}\\
      +1.09 \end{array}}$
      & $\scriptsize{\begin{array}[t]{c}\left(\begin{array}{rrr} +1.13 &     +0.12 & \ms\+0.00 \\     -0.12 & +1.13 & \ms\+0.00 \\ \ms\+0.00 & \ms\+0.00 & +1.02 \end{array}\right)\vspace{4pt}\\
      \left[ \begin{array}{rrr} +1.13 &+1.13 &+1.02  \end{array}\right] \vspace{2pt}\\
      +1.09 \end{array}}$
      & $\scriptsize{\begin{array}[t]{c}\left(\begin{array}{rrr} -1.72 &     -0.48 &     +0.23 \\     -0.48 & -1.72 &     +0.23 \\     +0.15 &     +0.15 & -1.65 \end{array}\right)\vspace{4pt}\\
      \left[ \begin{array}{rrr} -1.25 &-2.31 &-1.54  \end{array}\right] \vspace{2pt}\\
      -1.70 \end{array}}\vspace{6pt}$
      \\
CZ-5  & $\scriptsize{\begin{array}[t]{c}\left(\begin{array}{rrr} +3.12 & \ms\+0.00 & \ms\+0.00 \\ \ms\+0.00 & +3.12 & \ms\+0.00 \\ \ms\+0.00 & \ms\+0.00 & +3.22 \end{array}\right)\vspace{4pt}\\
      \left[ \begin{array}{rrr} +3.12 &+3.12 &+3.22  \end{array}\right] \vspace{2pt}\\
      +3.15 \end{array}}$
      & $\scriptsize{\begin{array}[t]{c}\left(\begin{array}{rrr} +1.89 & \ms\+0.00 & \ms\+0.00 \\ \ms\+0.00 & +1.89 & \ms\+0.00 \\ \ms\+0.00 & \ms\+0.00 & +1.56 \end{array}\right)\vspace{4pt}\\
      \left[ \begin{array}{rrr} +1.89 &+1.89 &1.56  \end{array}\right] \vspace{2pt}\\
      +1.78 \end{array}}$
      & $\scriptsize{\begin{array}[t]{c}\left(\begin{array}{rrr} +1.01 &     +0.02 & \ms\+0.00 \\     -0.02 & +1.01 & \ms\+0.00 \\ \ms\+0.00 & \ms\+0.00 & +0.77 \end{array}\right)\vspace{4pt}\\
      \left[ \begin{array}{rrr} +1.01 &+1.01 &+0.77  \end{array}\right] \vspace{2pt}\\
      +0.93 \end{array}}$
      & $\scriptsize{\begin{array}[t]{c}\left(\begin{array}{rrr} +1.01 &     +0.25 & \ms\+0.00 \\     -0.25 & +1.01 & \ms\+0.00 \\ \ms\+0.00 & \ms\+0.00 & +0.77 \end{array}\right)\vspace{4pt}\\
      \left[ \begin{array}{rrr} +1.01 &+1.01 &+0.77  \end{array}\right] \vspace{2pt}\\
      0.93 \end{array}}$
      & $\scriptsize{\begin{array}[t]{c}\left(\begin{array}{rrr} -1.76 &     -0.65 &     +0.22 \\     -0.65 & -1.76 &     +0.22 \\     +0.02 &     +0.02 & -1.58 \end{array}\right)\vspace{4pt}\\
      \left[ \begin{array}{rrr} -1.11 &-2.44 &-1.55  \end{array}\right] \vspace{2pt}\\
      -1.70 \end{array}}\vspace{6pt}$
      \\
CZ-6  & $\scriptsize{\begin{array}[t]{c}\left(\begin{array}{rrr} +3.12 & \ms\+0.00 & \ms\+0.00 \\ \ms\+0.00 & +3.12 & \ms\+0.00 \\ \ms\+0.00 & \ms\+0.00 & +3.20  \end{array}\right)\vspace{4pt}\\
      \left[ \begin{array}{rrr} +3.12 &+3.12 &+3.20  \end{array}\right] \vspace{2pt}\\
      +3.15 \end{array}}$
      & $\scriptsize{\begin{array}[t]{c}\left(\begin{array}{rrr} +2.09 & \ms\+0.00 & \ms\+0.00 \\ \ms\+0.00 & +2.09 & \ms\+0.00 \\ \ms\+0.00 & \ms\+0.00 & +1.68 \end{array}\right)\vspace{4pt}\\
      \left[ \begin{array}{rrr} +2.09 &+2.09 &+1.68  \end{array}\right] \vspace{2pt}\\
      +1.95 \end{array}}$
      & $\scriptsize{\begin{array}[t]{c}\left(\begin{array}{rrr} +0.97 &     +0.35 & \ms\+0.00 \\     -0.35 & +0.97 & \ms\+0.00 \\ \ms\+0.00 & \ms\+0.00 & +0.61 \end{array}\right)\vspace{4pt}\\
      \left[ \begin{array}{rrr} +0.97 &+0.97 &+0.61  \end{array}\right] \vspace{2pt}\\
      +0.85 \end{array}}$
      & $\scriptsize{\begin{array}[t]{c}\left(\begin{array}{rrr} +0.97 &     +0.35 & \ms\+0.00 \\     -0.35 & +0.97 & \ms\+0.00 \\ \ms\+0.00 & \ms\+0.00 & +0.61 \end{array}\right)\vspace{4pt}\\
      \left[ \begin{array}{rrr} +0.97 &+0.97 &+0.61  \end{array}\right] \vspace{2pt}\\
      +0.85 \end{array}}$
      & $\scriptsize{\begin{array}[t]{c}\left(\begin{array}{rrr} -1.79 &     -0.77 &     +0.22 \\     -0.77 & -1.79 &     +0.22 \\     -0.05 &     -0.05 & -1.53 \end{array}\right)\vspace{4pt}\\
      \left[ \begin{array}{rrr} -1.02 &-2.57 &-1.51  \end{array}\right] \vspace{2pt}\\
      -1.70 \end{array}}\vspace{6pt}$
      \\      
ST    & $\scriptsize{\begin{array}[t]{c}\left(\begin{array}{rrr} +3.13 & \ms\+0.00 & \ms\+0.00 \\ \ms\+0.00 & +3.13 & \ms\+0.00 \\ \ms\+0.00 & \ms\+0.00 & +3.19 \end{array}\right)\vspace{4pt}\\
      \left[ \begin{array}{rrr} +3.13 &+3.13 &+3.19  \end{array}\right] \vspace{2pt}\\
      +3.15 \end{array}}$
      & $\scriptsize{\begin{array}[t]{c}\left(\begin{array}{rrr} +2.22 & \ms\+0.00 & \ms\+0.00 \\ \ms\+0.00 & +2.22 & \ms\+0.00 \\ \ms\+0.00 & \ms\+0.00 & +1.75 \end{array}\right)\vspace{4pt}\\
      \left[ \begin{array}{rrr} +2.22 &+2.22 &+1.75  \end{array}\right] \vspace{2pt}\\
      +2.06 \end{array}}$
      & $\scriptsize{\begin{array}[t]{c}\left(\begin{array}{rrr} +0.96 &     +0.42 & \ms\+0.00 \\     -0.42 & +0.96 &     +0.50 \\ \ms\+0.00 & \ms\+0.00 & +0.50 \end{array}\right)\vspace{4pt}\\
      \left[ \begin{array}{rrr} +0.96 &+0.96 &+0.50  \end{array}\right] \vspace{2pt}\\
      +0.81 \end{array}}$
      & $\scriptsize{\begin{array}[t]{c}\left(\begin{array}{rrr} +0.96 &     +0.42 & \ms\+0.00 \\     -0.42 & +0.96 & \ms\+0.00 \\ \ms\+0.00 & \ms\+0.00 & +0.50 \end{array}\right)\vspace{4pt}\\
      \left[ \begin{array}{rrr} +0.96 &+0.96 &+0.50  \end{array}\right] \vspace{2pt}\\
      +0.81 \end{array}}$
      & $\scriptsize{\begin{array}[t]{c}\left(\begin{array}{rrr} -1.82 &     -0.86 &     +0.22 \\     -0.86 & -1.82 &     +0.22 \\     -0.09 &     -0.09 & -1.48 \end{array}\right)\vspace{4pt}\\
      \left[ \begin{array}{rrr} -0.96 &-2.69 &-1.48  \end{array}\right] \vspace{2pt}\\
      -1.71 \end{array}}\vspace{6pt}$
      \\
\end{tabular}
\end{ruledtabular}
\end{table*}

In Sec.~\ref{sec:amodes-evolution}, we have observed that the projection of intensities on the $\tilde{A}''$ mode ($\alpha$) in combination with the bond polarizability parameters within the context of BPM are two important factors to consider in order to understand the change in intensities as well as their inversion with the increase in disorder. Here, in order to have a microscopic understanding of changes in Raman intensity of $A''$ and $A'''$ with respect to atomic occupation in the different Wyckoff positions, we investigate the anomalous Born effective charges. 
The latter are the deviation of the mean value obtained from the BEC eigenvalues in 2$a$, 2$b$, 2$c$, 2$d$, and 8$g$ Wyckoff sites from the nominal charge on the site,  $\Delta Z^{*} = \sum_{i=1}^{3} Z_{i}/3 - Z^{*}_\mathrm{nominal}$, where $Z_{i}$ correspond to the eigenvalues of the BEC tensor. Raman activity is usually higher when large changes in polarizability are observed. In that context, deviation from the nominal charges on atomic sites can be valuable to understand how the charge redistribution during mode vibration can affect changes in Raman intensity.

\begin{figure}[h]
\centering
\includegraphics{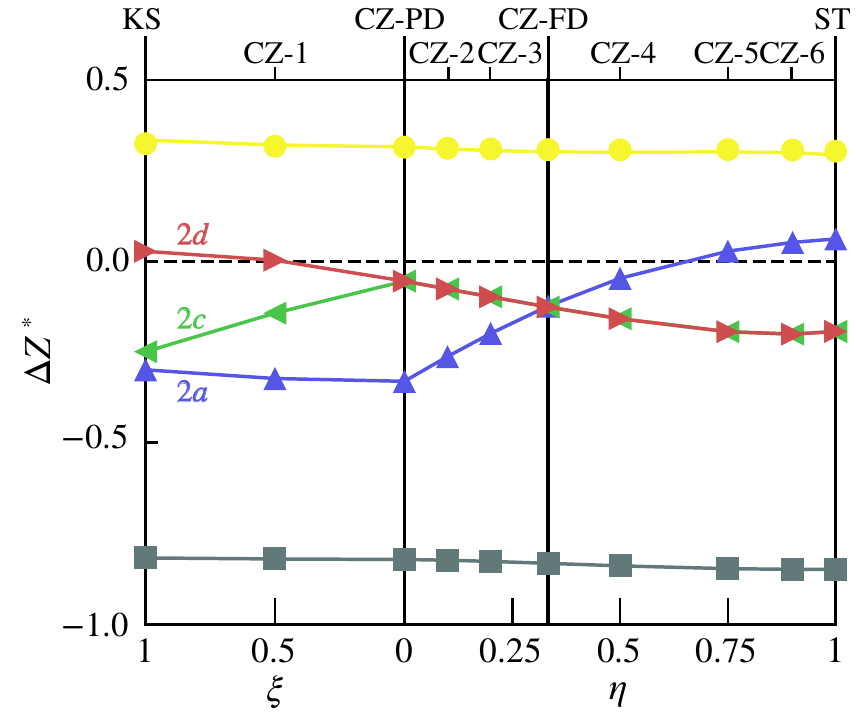}
\caption{Evolution of the anomalous Born effective charges of the different atoms (Sn and S atoms are in grey and yellow respectively, while the atoms at the 2$a$, 2$c$, and 2$d$ sites are in blue, green, and red) as a function of disorder from KS through the intermediate phases until ST.
}
\label{fig:anomalous-bec}

\end{figure}

The evolution the anomalous BEC is shown in Fig.~\ref{fig:anomalous-bec}.
At first, we observe that in the extreme ends corresponding KS and ST phases, $\Delta Z^{*}$ has almost the same value at the 2$c$ site while the 2$a$ and 2$d$ switch their corresponding values going from KS to ST. As noted previously, since the polarizability for bonds involving Cu is higher than for those involving Zn, for ST, with 2$c$ and 2$d$ sites occupied by Cu showing a larger $\Delta Z^{*}$ would tend to have a lower intensity for the $A'''$ mode compared to $A''$ based on the equations in BPM. On the other hand, for KS, while the $\Delta Z^{*}$ is close to zero for the 2$d$ site, there is significant deviation at 2$a$ and 2$c$ sites which are occupied by Cu and Zn respectively. Again, since bonds involving Cu have a larger polarizability parameter compared to those involving Zn, the $A''$ mode probably has large changes in polarizability and hence a higher intensity than $A'''$, whereas the combined effect of 2$a$ and 2$c$ site (based on BPM) for $A'''$ mode leads to a relatively smaller intensity. We will understand the intensity changes for the intermediate phases using  similar arguments based on the polarizability parameters of the bonds involving different atomic occupation at the Wyckoff site and $\Delta Z^{*}$. Between KS to CZ-PD, $\Delta Z^{*}$ decreases slightly at 2$a$, and since it is occupied by Cu, the intensity of $A''$ mode decreases as well while the combination of decrease (resp. increase) of $\Delta Z^{*}$ in the 2$a$/2$d$ (resp. 2$c$) only decreases the intensity of $A'''$ mode slightly. Beyond CZ-PD and until CZ-FD, $\Delta Z^{*}$ increases for 2$a$, while we would expect the intensity to increase, it decreases further because now the occupation of Zn (with smaller polarizability) is increasing. On the other hand, the changes in $\Delta Z^{*}$ are compensated by the decrease (resp. increase) in 2$c$/2$d$ (resp. 2$a$) sites still not showing significant intensity change for $A'''$. At CZ-FD, $\Delta Z^{*}$ is the same in the 2$a$, 2$c$, and 2$d$ Wyckoff sites indicating an equal redistribution of charges between them. After CZ-FD, while we would expect $A'''$ to decrease as the Cu occupation increases in the 2$c$ and 2$d$ sites, the movement of $\Delta Z^{*}$ at the 2$a$ to a positive value counteracts this trend, and hence starts to increase. In the same way, increase in $\Delta Z^{*}$ to a positive value despite increasing occupation of Zn, increases the intensity of $A''$ mode until ST.

\twocolumngrid
\bibliography{paper.bib}
\bibliographystyle{apsrev4-1}
\end{document}